\documentclass[8.5pt,twoside,twocolumn]{article}
\usepackage[super,sort&compress,comma]{natbib} 
\usepackage{mhchem}
\usepackage{times,mathptmx}
\usepackage{times}
\usepackage{sectsty}
\usepackage{balance} 
\usepackage{color}
\usepackage{rotating}
\usepackage{multirow}
\usepackage{graphicx} 
\usepackage{lastpage}
\usepackage[format=plain,justification=raggedright,singlelinecheck=false,font=small,labelfont=bf,labelsep=space]{caption} 
\usepackage{fancyhdr}
\begin{document}

\thispagestyle{plain}
\fancypagestyle{plain}{
\renewcommand{\headrulewidth}{1pt}}
\renewcommand{\thefootnote}{\fnsymbol{footnote}}
\renewcommand\footnoterule{\vspace*{1pt}%
\hrule width 3.4in height 0.4pt \vspace*{5pt}} 
\setcounter{secnumdepth}{5}

\makeatletter 
\def\subsubsection{\@startsection{subsubsection}{3}{10pt}{-1.25ex plus -1ex minus -.1ex}{0ex plus 0ex}{\normalsize\bf}} 
\def\paragraph{\@startsection{paragraph}{4}{10pt}{-1.25ex plus -1ex minus -.1ex}{0ex plus 0ex}{\normalsize\textit}} 
\renewcommand\@biblabel[1]{#1}            
\renewcommand\@makefntext[1]%
{\noindent\makebox[0pt][r]{\@thefnmark\,}#1}
\makeatother 
\renewcommand{\figurename}{\small{Fig.}~}
\sectionfont{\large}
\subsectionfont{\normalsize} 


\twocolumn[
  \begin{@twocolumnfalse}
\noindent\LARGE{\textbf{Unravelling the quantum-entanglement effect of noble gas coordination on the spin ground state of CUO$^\dag$}}
\vspace{0.6cm}

\noindent\large{\textbf{Pawe{\l} Tecmer,$^{\ast}$\textit{$^{a\ddag}$} Katharina Boguslawski,$^{\ast}$\textit{$^{a\ddag}$} \"Ors Legeza,\textit{$^{b}$} and Markus Reiher\textit{$^{a}$}}}\vspace{0.5cm}


\noindent \textbf{\small{DOI: 10.1039/C3CP53975J}}
\vspace{0.6cm}

\noindent \normalsize{The accurate description of the complexation of the CUO molecule by Ne and Ar noble gas matrices represents a challenging task for present-day quantum chemistry. Especially, the accurate prediction of the spin ground state of different CUO--noble-gas complexes remains elusive. In this work, the interaction of the CUO unit with the surrounding noble gas matrices is investigated in terms of complexation energies and dissected into its molecular orbital quantum entanglement patterns. Our analysis elucidates the anticipated singlet--triplet ground-state reversal of the CUO molecule diluted in different noble gas matrices and demonstrates that the strongest uranium-noble gas interaction is found for CUOAr$_4$ in its triplet configuration.}
\vspace{0.5cm}
 \end{@twocolumnfalse}
]

\section{Introduction}


\footnotetext{\dag~Electronic Supplementary Information (ESI) available: Electronic energies and additional entanglement diagrams. See DOI: 10.1039/b000000x/}
\footnotetext{\textit{$^{a}$~ETH Z\"urich, Laboratory of Physical Chemistry, Wolfgang-Pauli-Str. 10, CH-8093 Z\"urich, Switzerland. tecmer@mcmaster.ca. bogusl@mcmaster.ca}}
\footnotetext{\textit{$^{b}$~Wigner Research Center for Physics, Strongly Correlated Systems "Lend\"ulet" Research Group, H-1525 Budapest, Hungary.}}

\footnotetext{\ddag~Present address: Department of Chemistry and Chemical Biology, McMaster University, Hamilton, Ontario, Canada.}





Uranium chemistry represents a diverse field of chemistry \cite{bursten_91,denning07,Actinides_bible,MRS_bulletin}. 
Apart from the employment of large uranium complexes in nuclear waste reprocessing \cite{PUREX1,sood95,PUREX2} 
and catalysis \cite{U_catalysis_cc,U_catalysis_n,Meyer_nature_2008,Meyer_book_2008}, 
there exists a number of small uranium-containing compounds that exhibit peculiar 
and unusual features \cite{Castro-Rodriguez2004,Hayton2005,Gary_2008, Groenewold_2010,CUO_Chem_Sci}. 
Examples are the notorious quintuple bond for the U$_2$ 
dimer \cite{Roos_U2} and the discovery of the unexpected complexation of uranium compounds by noble gases \cite{wang_2004,Heaven_UO2_04,ivan_2010}. 

Especially, the remarkable and even "mysterious" interaction of the CUO molecule with different noble gas matrices has attracted much attention of both experimentalists and quantum chemists over the past decade. In experimental studies, dilution of CUO into different noble gas environments led to a blue shift of the characteristic asymmetric UO and UC vibrational frequencies (by $\sim$70 and $\sim$200 cm$^{-1}$, respectively) when the noble gas surrounding was systematically varied from Ne to Ar \cite{CUO_93,CUO_Nb_science}. Based on these results, a ground state spin change from the $^1\Sigma^+$ singlet to the $^3\Phi$ triplet state of the CUO unit has been anticipated if the noble gas matrix is altered from Ne to Ar \cite{CUO_93,CUO_Nb_science,CUO_Nb_JACS,CUO_CEJ,CUO_Nb_JACSa,CUO_Nb_IC}. So far, this hypothesis could not be confirmed in any quantum chemical study.

Due to partially occupied $f$-orbitals and the high nuclear charge number of the U atom, a balanced treatment of both relativistic \cite{Timo_overview,Trond_rev_2012} and electron correlation effects is essential \cite{CUO_2000,Bursten_2003,CUO_Laura,CUO_Ivan,CUO_CI,CUO_pawel}, which remains a challenging task even 
for present-day quantum chemistry \cite{actinoid_rev_2012,pawel_saldien}. 
{In general, an accurate description of uranium-containing compounds requires a four-component fully relativistic framework, where in the case of linear molecules, the quantum number of the projected total angular momentum  $\Omega$ is a good quantum number.}
Since, however, spin--orbit coupling in the CUO molecule is small compared to the correlation energy \cite{CUO_Laura,CUO_Ivan,CUO_pawel}, it is sufficient to consider scalar relativistic effects only 
for a qualitative analysis, while spin--orbit coupling may be added \emph{a posteriori} 
in a perturbative treatment \cite{Malmqvist2002}. 
It is then most fortunate that, within such a scalar-relativistic description, electronic states may be still classified according to spin and projected orbital angular momentum symmetries as $^1\Sigma$, $^3\Phi$, and so forth. 

Yet, even in scalar relativistic calculations, the proper prediction of spin-state energetics for the bare CUO molecule remains extremely challenging. 
It was shown by one of us that (all-electron) density functional theory (DFT) predicts the $^3\Phi$ state to be slightly lower in energy than the $^1\Sigma^+$ state \cite{CUO_pawel}. 
Similarly, Hartree--Fock calculations yield a triplet ground state \cite{CUO_pawel} which may limit the applicability of single-reference methods relying on a Hartree--Fock reference wave function. 
Besides, the anticipated singlet--triplet spin crossover may have a significant multi-determinant character and thus \emph{ab initio} multi-reference wave function approaches are required. 
One possibility would be the application of the standard complete active space self-consistent field (CASSCF) method \cite{Werner_1985}, 
but it is only applicable to the bare CUO unit since the active space necessary to 
describe noble-gas--CUO complexes would exceed the current limit of the method of  
say, 18 correlated electrons in 18 spatial orbitals.

An alternative ansatz, which allows one to consider much larger active spaces than CASSCF, is the density matrix renormalization group (DMRG) algorithm \cite{scholl05} developed by White\cite{white} for solid state physics. 
The quantum chemical extension of DMRG \cite{ors_springer,marti2010b,chanreview} has been successfully applied in many areas of chemistry, including very challenging systems such us open--shell transition metal complexes \cite{marti2008,fenoDMRG,kurashige2013}. 
An advantage of DMRG is that it allows to capture all types of electron correlation effects
(dynamic, static and non-dynamic) in a given active space in a balanced way \cite{entanglement_letter}. 
In other words, the DMRG wave function is rather flexible to adjust to all changes 
in electron correlation induced by structural changes \cite{orbitalordering} such as ligand coordination. 

In this work, we present a DMRG study of CUO, CUONe$_4$ and CUOAr$_4$ in their singlet and 
triplet states. 
An entanglement analysis as outlined in Refs.~\citenum{entanglement_letter,entanglement_bonding_2013} 
is employed to dissect the origin of the stabilization
of CUO in different noble gas matrices in terms of orbital correlations.
This analysis will allow us to elucidate the singlet-triplet state reversal of the CUO molecule when the noble gas environment is varied. 

\section{\large Computational details}

\subsection{Basis sets and relativity.}
%
For the C, O and U atom, a contracted triple-$\zeta$ polarization (TZP) ANO-RCC basis sets was used: (14$s$9$p$4$d$3$f$2$g)\rightarrow[4s$3$p$2$d$1$f$]
\cite{ANO-RCC_main_group} for C and O, and (26$s$23$p$17$d$13$f$5$g$3$h)\rightarrow[9s$8$p$6$d$2$f$] \cite{ANO-RCC_U} for U. For the noble gas atoms (Ne and Ar), a contracted double-$\zeta$ polarization (DZP) ANO-RCC basis sets was employed: (14$s$9$p$4$d$3$f$2$g)\rightarrow[3s$2$p$1$d$]\cite{ANO-RCC_main_group}.

Scalar relativistic effects were incorporated through the 10th-order Douglas--Kroll--Hess (DKH10) Hamiltonian \cite{dkh2,Reiher_2004a,Reiher_2004b} as implemented in the \textsc{MOLPRO 2010.1} quantum chemical package. \cite{MOLPRO}

{The value of spin--orbit coupling has been extracted from the Fock-Space coupled cluster singles and doubles results with and without spin-orbit coupling provided in Ref.~\citenum{CUO_pawel}. In particular, the values of 0.40 eV and 0.33 eV were assigned to $\Omega=2$ and $\Omega=3$ states of the $^3\Phi$ CUO, respectively. These energy contributions were added \emph{a posteriori} to the scalar relativistic results.}

\subsection{CASSCF.}
%
All CASSCF calculations were performed with the \textsc{MOLPRO 2010.1} quantum chemical package \cite{MOLPRO} imposing C$_{2v}$ point group symmetry.
For $^1\Sigma^{+}$ CUO, $^1$A$_1 $ CUONe$_4$ and $^1$A$_1 $ CUOAr$_4$ and all U--Ng distances,
the active spaces comprise 12 electrons in 12 orbitals (CAS(12,12)SCF). Such an active space
contains the bonding and antibonding combinations of the 2$p_{\rm z}$-orbitals of C and O
with the U 6$d$- and 5$f$-orbitals (4 orbitals in A$_1$ symmetry) as well as the bonding and
antibonding combinations of the 2$p_{\rm x}$- and 2$p_{\rm y}$- orbitals of C and O with 
the U 6$d$- and 5$f$-orbitals (4 orbitals in B$_1$ and B$_2$ symmetry, respectively).
For the corresponding triplet states, the nonbonding 5$f_{\phi}$-orbital of the U atom
was additionally included in the active space, which results in CAS(12,14)SCF calculations
(one additional orbital in B$_1$ and one in B$_2$ symmetry). 
It is well-known that such nonbonding orbitals contribute very little to the total correlation energy in the CASSCF approach~\cite{pierloot05}, and therefore CAS(12,12)SCF can be compared to CAS(12,14)SCF. 
Note that no noble gas orbitals
are contained in any of the CASSCF active spaces.

\subsection{DMRG.}
All DMRG calculations were performed with the \textsc{Budapest DMRG} program \cite{dmrg_ors}.
As orbital
basis, the natural orbitals obtained from CASSCF calculations described above were taken. 
The active spaces were extended to CAS(14,40) for the bare CUO molecule and to CAS(38,36) for all
CUONg$_4$ (Ng = Ne, Ar) complexes and U--Ng distances. 
In particular, for the CUONg$_4$ molecules, 4 occupied and 7 unoccupied orbitals were
added in A$_1$ symmetry, 3 occupied and 2 unoccupied in B$_1$ and B$_2$ symmetry,
respectively, and 3 occupied and 2 unoccupied in A$_2$ symmetry. 
More detailed information concerning molecular orbitals used in our DMRG calculation, that is, their type and their main atomic contributions, can be found in Table~\ref{tbl:DMRG_orbitals}.

To enhance DMRG convergence, the orbital ordering was optimized as described in Ref.~\citenum{orbitalordering} and the number of renormalized active-system states
$m$ was chosen
dynamically according to a predefined threshold value for the quantum information loss
\cite{legeza_dbss} employing the dynamic block state selection approach \cite{legeza_dbss2,legeza_dbss3}. As initial guess, the
dynamically-extended-active-space procedure was applied \cite{legeza_dbss}.
In the DMRG calculations, the maximum number of renormalized
active-system states $m_{\rm max}$ was varied from 1024 to 2048, while the minimum number $m_{\rm min}$
was set to 512 if not stated otherwise. To avoid local minima, the minimum number of renormalized active-system states used during the initialization procedure $m_{\rm start}$ was set equal to $m_{\rm max}$. The quantum information loss was chosen to be $10^{-5}$ in all calculations. 

\begin{table*}[t]
\centering
     \caption{Numbering scheme and atomic contributions for the molecular orbitals of CUONe$_4$ and CUOAr$_4$}
     \label{tbl:DMRG_orbitals}
\begin{tabular}{l| cl| l| cl} 
\hline
\hline
\multicolumn{1}{c|}{Orbital} &   
\multicolumn{2}{c}{CUONe$_4$} &
\multicolumn{1}{|c|}{Orbital} & 
\multicolumn{2}{c}{CUOAr$_4$} \\
\cline{2-3}
\cline{5-6}
index& Mol. orb. & Atomic contrib. & index & Mol. orb. & Atom. contrib.\\
\hline
\hline
1    &A$_1$&Ne $[2p_{x,y}]$               & 1    &A$_1$&Ar  $[3p_{x,y}]$               \\
2    &A$_1$&Ne $[2p_{x,y}]$               & 2    &A$_1$&Ar  $[3p_{x,y}]$               \\
3    &A$_1$&Ne $[2p_{z}]$                 & 3    &A$_1$&Ar  $[3p_{z}]$                 \\
4    &$\sigma$&C/O $[2_{s}]$                 & 4    &$\sigma$&C/O $[2_{s}]$                 \\
5    &$\sigma$&C/O $[2p_{z}]$                & 5    &$\sigma$&C/O $[2p_{z}]$                \\
6    &$\sigma$&C/O $[2p_{z}]$                & 6    &$\sigma$&C/O $[2p_{z}]$                \\
7    &$\sigma$&C/O $[2p_{z}]$                & 7    &$\sigma$&C/O $[2p_{z}]$                \\
8    &$\sigma$&C/O $[2p_{z}]$                & 8    &$\sigma$&C/O $[2p_{z}]$                \\
9/35 &$\delta$&U $[6d_{xy, x^2 -y^2}]$       & 9/35 &$\delta$&U $[6d_{xy, x^2 -y^2}]$       \\
10   &$\sigma$&U $[6_s]$                     & 10   &$\sigma$&U $[6_s]$                     \\
11   &$\sigma$&U $[7p_{z}]$                  & 11   &$\sigma$&U $[7p_{z}]$                  \\
12   &$\sigma$&U $[4f_{z^3}]$                & 12   &$\sigma$&U $[4f_{z^3}]$                \\
13/36&$\delta$&U $[5f_{xyz,z(x^2-y^2)}]$     & 13/36&$\delta$&U $[5f_{xyz,z(x^2-y^2)}]$     \\
14/23&B$_1$/B$_2$   &Ne $[2p_x]$                   & 14/23&B$_1$/B$_2$   &Ar $[3p_x]$                   \\
15/24&B$_1$/B$_2$   &Ne $[2p_x]$                   & 15/24&B$_1$/B$_2$   &Ar $[3p_x]$                   \\
16/25&B$_1$/B$_2$   &Ne $[2p_x]$                   & 16/25&B$_1$/B$_2$   &Ar $[3p_x]$                   \\
17/26&$\pi$   &C/O$[2p_{x,y}]$               & 17/26&$\pi$   &C/O$[2p_{x,y}]$               \\
18/27&$\pi$   &C/O$[2p_{x,y}]$               & 18/27&$\pi$   &C/O$[2p_{x,y}]$               \\
19/28&$\pi$   &C/O$[2p_{x,y}]$               & 19/28&$\pi$   &C/O$[2p_{x,y}]$               \\
20/29&$\pi$   &C/O$[2p_{x,y}]$               & 20/29&$\pi$   &C/O$[2p_{x,y}]$               \\
21/30&$\phi$  &U$[5f_{x^3-3xy^2, 3x^2y-y^3}]$& 21/30&$\phi$  &U$[5f_{x^3-3xy^2, 3x^2y-y^3}]$\\
22/31&$\pi$   &U $[7p_x]$                    & 22/31&$\pi$   &U $[7p_x]$                    \\
32   &A$_1$&Ne $[2p_{x,y}]$               & 32   &A$_1$&Ar $[3p_{x,y}]$               \\
33   &A$_1$&Ne $[2p_{x,y}]$               & 33   &A$_1$&Ar $[3p_{x,y}]$               \\
34   &A$_1$&Ne $[2p_{z}]$                 & 34   &A$_1$&Ar $[3p_{z}]$                 \\
\hline
\hline
\end{tabular}
\end{table*}

\begin{table*}[t]
\caption{Complexation energies $\rm{D_e}$ of four Ng atoms to the CUO moiety, 
U--Ng bond lengths $\rm{r_e}$, and dipole moments $\rm{DM}$ of CUONg$_4$ (Ng = Ne, Ar) 
obtained from CASSCF and DMRG calculations. The U--C and U--O distances in the 'Molecule' column 
are taken from Ref. \citenum{CUO_pawel}. $^3\Phi_{(\rm v)}$: $^3\Phi$ state for CUO$^{(\rm v)}$Ng$_4$. $^3\Phi_{(\rm a)}$: $^3\Phi$ state for CUO$^{(\rm a)}$Ng$_4$.}\label{tab:data}
{\small
\begin{center}
\begin{tabular*}{\textwidth}{@{\extracolsep{\fill}}llccccccc}\hline \hline
	 & & \multicolumn{3}{c}{CASSCF}  & & \multicolumn{3}{c}{DMRG} \\ \cline{3-5} \cline{7-9} 
\multicolumn{2}{c}{Molecule}   &  $\rm{D_e}$[kJ $\cdot$ mol$^{-1}$(eV)]  &  $\rm{r_e}$[\AA{}] & DM[D]
&&  $\rm{D_e}$[kJ $\cdot$ mol$^{-1}$(eV)]  &  $\rm{r_e}$[\AA{}] & DM[D] \\ \hline\\
\multirow{3}{*}{\begin{sideways}   CUONe$_4$ \end{sideways}} &
$^1\Sigma^+$ (d$_{\rm{UC}}$=1.734 \AA{}, d$_{\rm{UO}}$=1.782 \AA{})&0.7 (0.01) &4.314 &4.40 & &1.6 (0.02) & 4.224 &4.27\\
&
{$^3\Phi_{\rm(v)}$} (d$_{\rm{UC}}$=1.734 \AA{}, d$_{\rm{UO}}$=1.782 \AA{}) &{1.3 (0.01)} &{3.856} &{2.53} & &{2.7 (0.03)} &{3.713}&{2.25}\\
&
{$^3\Phi_{\rm (a)}$} (d$_{\rm{UC}}$=1.840 \AA{}, d$_{\rm{UO}}$=1.811 \AA{})&{1.6 (0.02)} &{3.744} \ &{2.60} & &{3.4 (0.04)} &{3.597}  &{2.31}\\ \\
\multirow{3}{*}{\begin{sideways}   CUOAr$_4$ \end{sideways}} &
$^1\Sigma^+$ (d$_{\rm{UC}}$=1.738 \AA{}, d$_{\rm{UO}}$=1.788 \AA{})&2.1 (0.02) & 4.193&3.21 && 4.1 (0.04)&4.120 &3.06\\
&{$^3\Phi_{\rm (v)}$} (d$_{\rm{UC}}$=1.738 \AA{}, d$_{\rm{UO}}$=1.788 \AA{})&{3.8 (0.04)} &{3.905} &{1.31} & &{7.0 (0.07)} &{3.790}&{1.01}\\
&{$^3\Phi_{\rm (a)}$} (d$_{\rm{UC}}$=1.845 \AA{}, d$_{\rm{UO}}$=1.815 \AA{})&{3.6 (0.04)} &{3.898} &{2.47} & &{6.9 (0.07)} &{3.798} &{2.19}\\\\ \hline \hline
\end{tabular*}
\end{center}
}
\end{table*}

\section{The spin ground state of the CUO molecule}\label{sec:CUO}
The electronic structure of the bare CUO molecule bears considerable similarity to its isoelectronic analogs, UO$_2^{2+}$, NUO$^+$, and NUN \cite{pyykko,CUO_2000,kaltsoyannis,matsika_2001,XUY,pawel1}, where the U 6$p$-, 5$f$- and 6$d$-orbitals interact with the 2$s$- and 2$p$-orbitals of the lighter elements entailing a stable linear structure \cite{denning07}. Yet, the energetically higher lying atomic orbitals of the C atom (in contrast to O and N) destabilize the CUO complex compared to the other isoelectronic species \cite{CUO_Nb_science}. It is now well-established that the CUO molecule features a $^1\Sigma^+$ ground-state which is very close in energy to a $^3\Phi$ excited state \cite{CUO_Ivan,CUO_CI,CUO_pawel}.
The latter involves electron transfer from the bonding $\sigma$-orbital of the C atom to the nonbonding $\phi$-orbital of the U atom resulting in a $\sigma^1 \phi^1$ electronic configuration. This electron transfer leads to a significant elongation and weakening of the U--C bond \cite{Bursten_2003} compared to the $^1\Sigma^+$ ground state. 

Our scalar-relativistic CAS(12,12)SCF calculation correctly predicts the $^1\Sigma^+$ state (${\rm r_{UC}}=1.773$ \AA~ and ${\rm r_{UO}}=1.779$ \AA \cite{CUO_pawel}) to be the ground state of the bare CUO molecule, which is separated by only 0.71 eV from the first adiabatically excited $^3\Phi$ state, determined from a CAS(12,14)SCF calculation (${\rm r_{UC}}=1.836$ \AA~ and ${\rm r_{UO}}=1.808$ \AA  \cite{CUO_pawel}). This singlet--triplet splitting reduces to 0.60 eV in our scalar relativistic DMRG(14,40) calculations.

\emph{A posteriori} addition of spin--orbit coupling on top of the lowest-lying triplet state further decreases the singlet--triplet gap to 0.31 and 0.20 eV for CASSCF and DMRG, respectively. 
It is worth to mention that the value of 0.40 eV (for $\Omega$=2) used in this article agrees well with the perturbative treatment of spin--orbit coupling of 0.36 eV determined by Roos \textit{et al.} \cite{CUO_Laura} 

Remarkably, this energy splitting is very prone to different noble gas surroundings and ground-state spin-crossover of the CUO moiety can be induced upon complexation of different noble gas atoms. Our spin--orbit corrected DMRG energy splitting of 0.20 eV for the $^3\Phi_2$ excited state with respect to the $^1\Sigma^+_0$ ground state is in line with results obtained from multi-reference spin--orbit configuration interaction singles and doubles calculations which predict an energy gap of 0.17 eV \cite{CUO_CI}. 

\section{Noble gas complexation to CUO}

\begin{figure}[ht]
\centering
\includegraphics[width=0.4\linewidth]{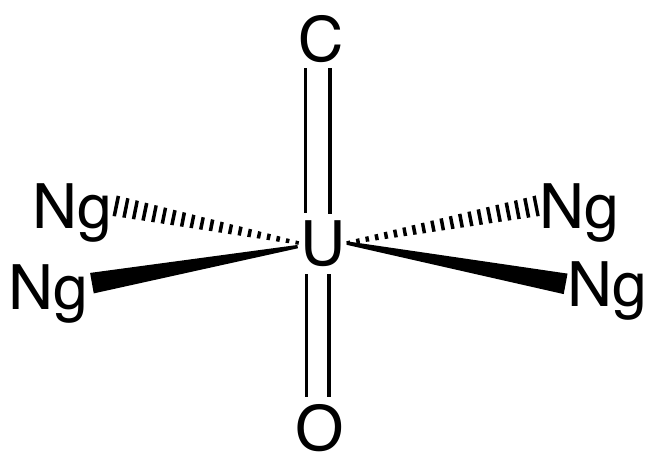}
\caption{Lewis structure of CUONg$_4$ for Ng = Ne, Ar.
}\label{fig:Structure}
\end{figure}

In this work, the noble gas environment is represented by four noble gas atoms arranged in an equatorial plane with respect to the CUO axis imposing C$_{\rm 4v}$ point group symmetry as depicted in Fig.~\ref{fig:Structure}. As discussed in 
Refs.~\citenum{CUO_Nb_JACS,CUO_CEJ,CUO_Nb_IC,CUO_pawel}, such a quadratic-planar coordination sphere constitutes a reliable model system for the extended noble gas matrix.  

{We choose the CUO geometries to be the same as in Ref.~34 which correspond to DFT optimized structures of the CUO and CUONg$_4$ molecules in their adiabatically and vertically excited triplet states and labeled them as CUO$^{(\rm v)}$Ng$_4$ and CUO$^{(\rm a)}$Ng$_4$, respectively.}
For the sake of simplicity and comparability to the bare CUO complex, we denote the singlet and triplet states of both CUONg$_4$ species (Ng = Ne, Ar) as $^1\Sigma^+$ and $^3\Phi$ (in accordance to C$_{\rm 4v}$ point group symmetry, the proper state labels are $^1$A$_{1}$ and $^3$E, respectively).

\subsection{Stabilization energies.}
\begin{figure*}[t]
\centering
\includegraphics[width=0.8\linewidth]{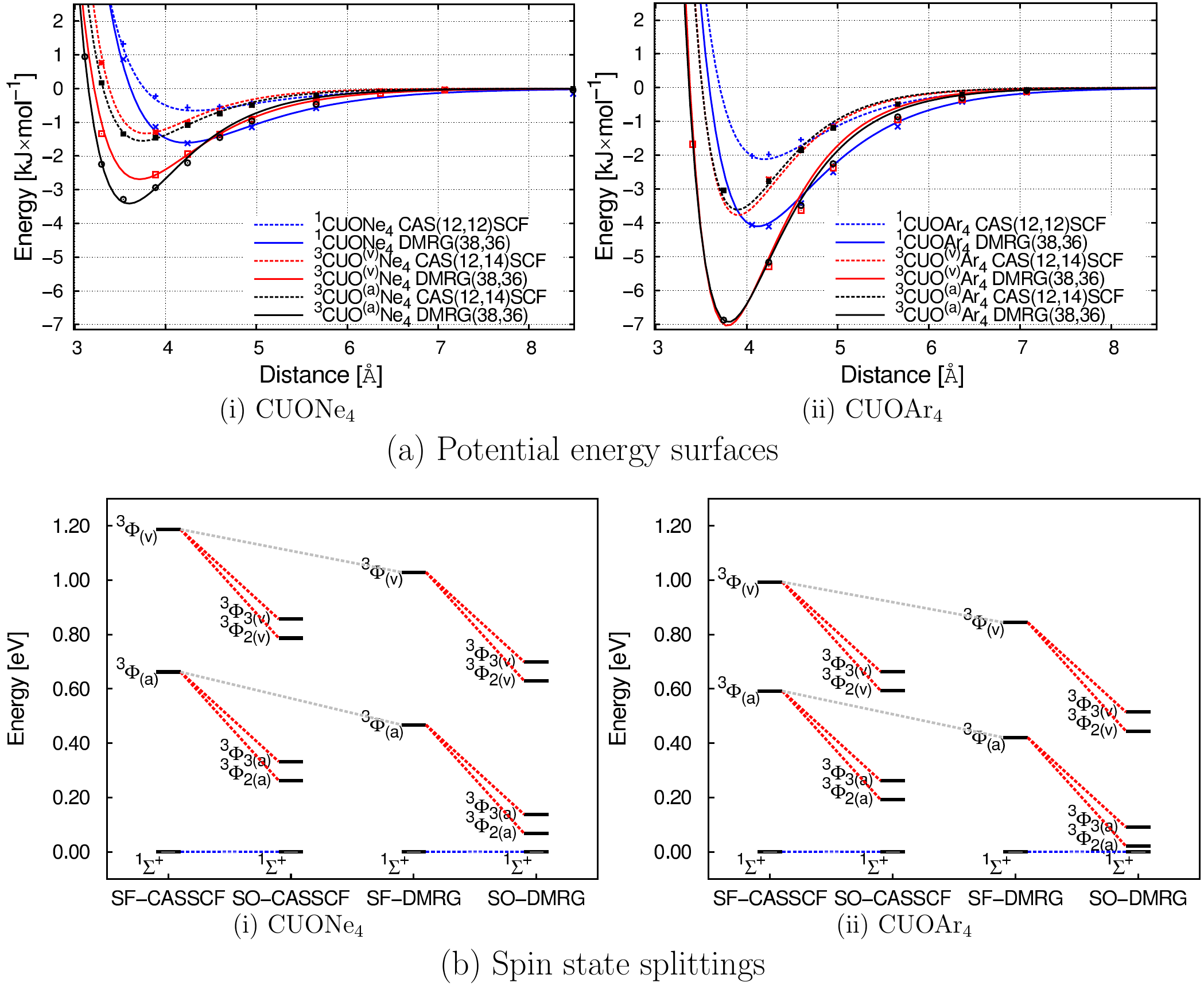}
\caption{Potential energy curves and spin-state splittings for CUONg$_4$ determined from DMRG and CASSCF calculations. (a) Reconstructed potential energy surfaces of (i) CUONe$_4$ and (ii) CUOAr$_4$ in kJmol$^{-1}$. The U--O and U--C distances are kept frozen (see Table.~\ref{tab:data}), while the U--Ne and U--Ar distances are varied. (b) Spin-free (SF) and spin--orbit (SO) corrected spin-state splittings of (i) CUONe$_4$ and (ii) CUOAr$_4$ in eV. $^3\Phi_{(\rm v)}$: $^3\Phi$ state for CUO$^{(\rm v)}$Ng$_4$. $^3\Phi_{(\rm a)}$: $^3\Phi$ state for CUO$^{(\rm a)}$Ng$_4$.
}\label{fig:PES-Ar}
\end{figure*}

The optimum distances between the Ng and the U atoms for all investigated electronic states are optimized 
by varying the U--Ng distance of all four noble gas atoms simultaneously (\emph{i.e.}, retaining C$_{\rm 4v}$ symmetry) in the range of 2.8 to 14.1 \AA{} (see the Supporting Information for further details), while the U--O and U--C bond lengths are kept frozen. As U--O and U--C bond distances, the values from Ref. \citenum{CUO_pawel} are taken (see Table~\ref{tab:data}). The potential energy curves obtained from CASSCF and DMRG
calculations are then fitted to a generalized Morse potential function \cite{Coxon_1992} and are plotted in Fig.~\ref{fig:PES-Ar}(a) for both CUONe$_4$ and CUOAr$_4$.

Exploring Fig.~\ref{fig:PES-Ar}(a), we observe an overall stabilization of the CUO molecule upon complexation of both Ng$_4$ surroundings. 
The complexation energy strongly depends on the spin state and on the specific Ng ligand. 
While the potential well depth is rather shallow for the Ne$_4$ matrix, it is twice as large in the Ar$_4$ environment for all electronic states investigated. In general, DMRG predicts a larger interaction energy between CUO and the noble gas atoms than CASSCF, which is more pronounced in the case of the Ar$_4$ than for the Ne$_4$ environment. 
 
Table \ref{tab:data} lists all complexation energies and dipole moments determined for the equilibrium U--Ng bond lengths for the singlet and the vertically and adiabatically excited triplet states of the CUONg$_4$ complexes. 
The complexation energy between CUO and Ng$_4$ is weakest in $^1\Sigma^+$ CUONe$_4$ (0.7 and 1.6 kJ$\cdot$mol$^{-1}$ for CASSCF and DMRG, respectively) and strongest in the vertically excited $^3\Phi$ state of CUOAr$_4$ (3.8 and 7.0 kJ$\cdot$mol$^{-1}$ for CASSCF and DMRG, respectively). 

We should mention that the DMRG stabilization energy of CUO by argon atoms is significantly smaller than the CASPT2 stabilization energy of UO$_2$ by argon atoms (7 \emph{vs.} 58  kJ$\cdot$mol$^{-1}$)~\cite{ivan_2010}. 

It is important to note that the interaction energy is similar for the vertically and the adiabatically excited states of CUOAr$_4$ (see Table~\ref{tab:data}). The shortest U--Ng bond length is found for $^3\Phi$ CUOAr$_4$, while the longest bond distance is observed for $^1\Sigma^+$ CUONe$_4$. 
For both Ng$_4$ environments, the equilibrium bond distances determined in DMRG calculations are generally shorter than for CASSCF. However, these differences are small ($\leq$ 0.15 \AA{}). Furthermore, CASSCF and DMRG yield similar dipole moments---although DMRG always provides smaller values, which overall agree well with previously reported theoretical data of 3.5D for the singlet and 2.4D for the triplet state in CUO \cite{CUO_Nb_science}. 
This observation indicates that changes in the dipole moment of the CUO unit are mainly affected by differences in the U--C and U--O bond lengths for the singlet and adiabatically excited triplet states rather than by complexation of noble gas atoms.

\begin{figure*}[t]
\centering
\includegraphics[width=0.75\linewidth]{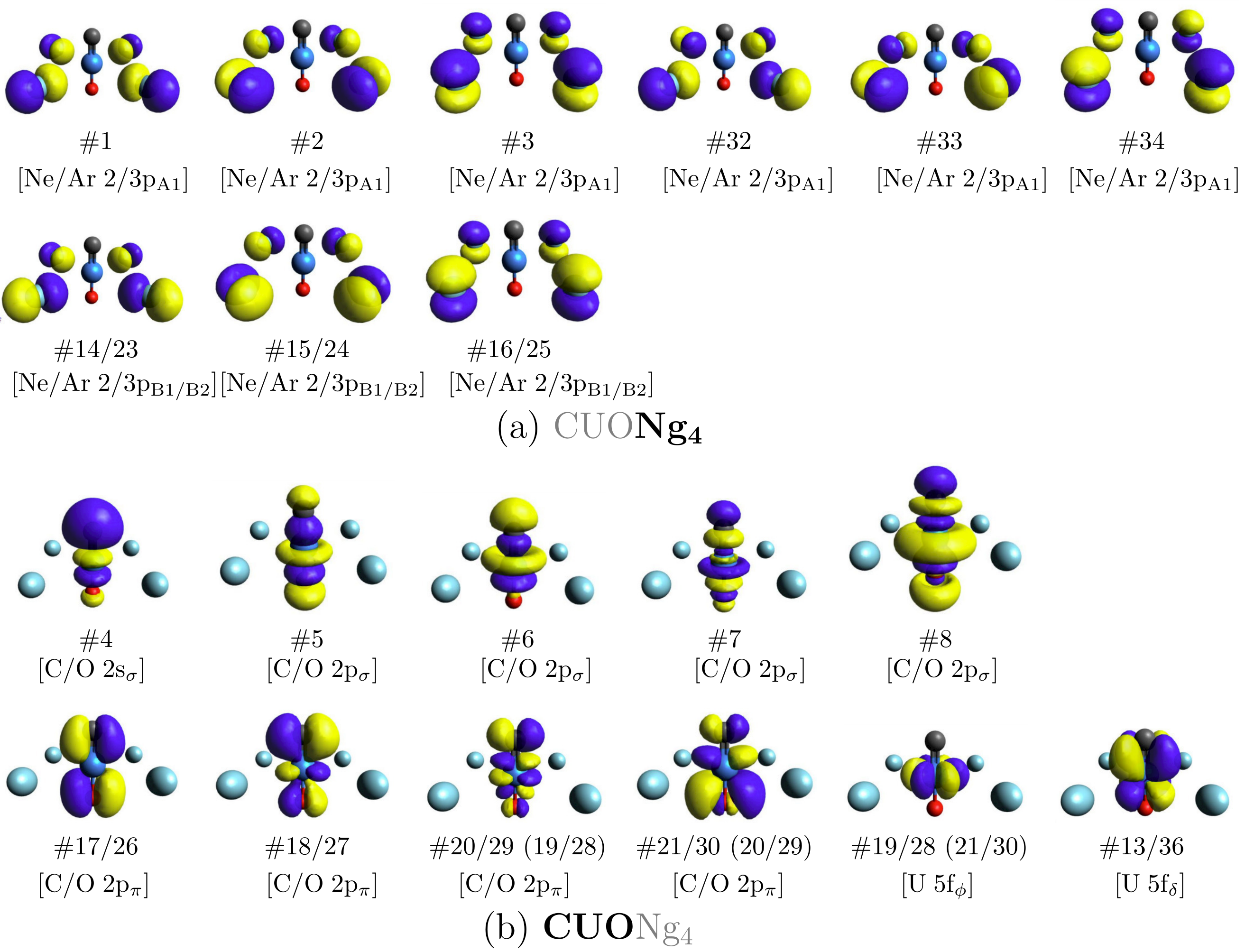}
\caption{
The most important 
molecular orbitals drawn with Avogadro program \cite{avogadro_1,avogadro_2}. The molecular orbitals are split into molecular orbitals centered on the noble gas surrounding in (a) and on the $^3$CUO$^{\rm (a)}$ unit in (b), respectively. The molecular orbitals are numbered according to their irreducible representation and CASSCF natural occupation number. Note that some orbital numbers differ for the singlet and triplet states. The corresponding molecular orbital indices for the singlet state are put in parentheses where required.
}\label{fig:orbitals}
\end{figure*}
Even though the interaction energy of CUO with the noble gas environment is small compared to the singlet--triplet splitting of the bare CUO molecule ($\leq$ 0.07 eV \emph{vs.} 0.20 eV), the complexation of Ng$_4$ to CUO considerably influences the CUO singlet--triplet gap. 
Fig.~\ref{fig:PES-Ar}(b) illustrates the changes of the spin-state splittings induced by the surrounding noble gases. 
Note that in Fig.~\ref{fig:PES-Ar}(b) all energies are measured with respect to the energy of the singlet 
state which was taken as reference point. 
As adiabatically excited states are lower in energy than the vertically excited states,  
the adiabatic energy difference yields the smallest singlet--triplet gaps. 
The spin-free CASSCF spin-state splitting of 0.66 eV in the Ne$_4$ surrounding is 
reduced to 0.59 eV in the Ar$_4$ environment. 
Similarly, the spin-free DMRG singlet--triplet gap of 0.47 eV determined for CUONe$_4$ decreases to 0.42 eV for CUOAr$_4$. 
A perturbative correction for spin--orbit coupling (energies taken from Ref. \citenum{CUO_pawel})
further lowers the $^3\Phi_{2({\rm a})}$ state to approach the $^1\Sigma_0^+$ state. 
For CASSCF, the energy gap of $^3\Phi_{2({\rm a})}$ and $^1\Sigma_0^+$ is lowered to 0.26 and 0.19 eV 
for CUONe$_4$ and CUOAr$_4$, 
respectively, while it 
reduces to 0.07 and 0.02 eV, respectively, in the DMRG calculations. 
In particular, an energy gap of 0.02 eV is below "chemical accuracy", which is of the order of 0.04 eV (or 4 kJ/mol), and hence the $^1\Sigma_{0}^+ $ and $^3\Phi_{2({\rm a})}$ states can be considered as energetically equivalent, where a thermal spin crossover \cite{M1,M2} ($^1$CUOAr$_4\, \leftrightarrow\, ^3$CUO$^{\rm (a)}$Ar$_4$) is possible. 

Note that a more rigorous treatment of weak interactions in these systems might further reduce the splitting or even reverse the states. 

\begin{figure*}[t]
\centering
\includegraphics[width=0.95\linewidth]{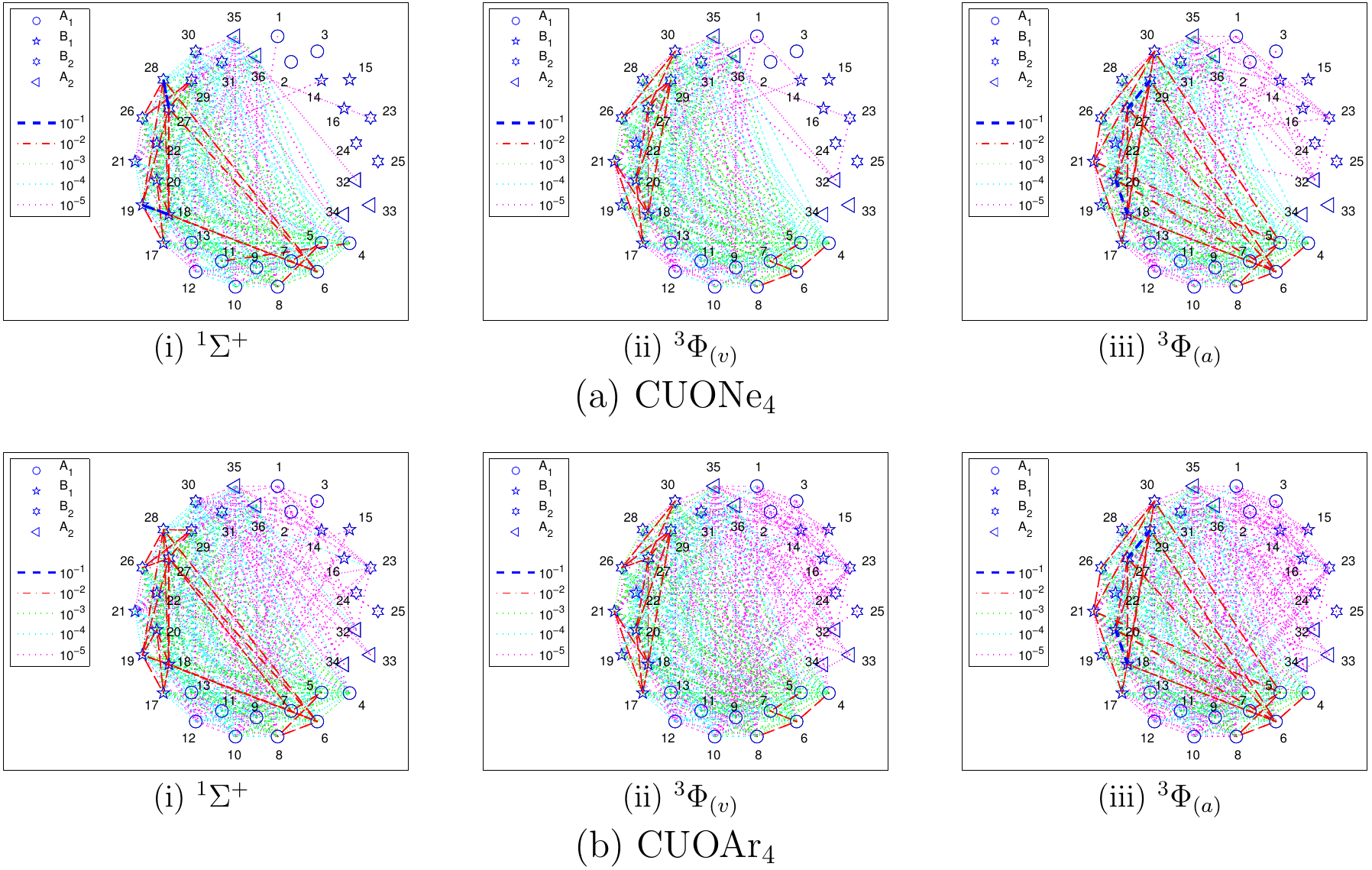}
\caption{
Quantum entanglement analysis of the CUO-Ng$_4$ (Ng = Ne, Ar) interaction. Mutual information 
of CUONg$_4$ orbital pairs determined at the equilibrium U--Ng distances. 
The first twelve orbitals (starting from \#1 and proceeding clockwise) represent Ng$_4$ molecular orbitals, while the remaining molecular orbitals are centered on the CUO moiety. The interaction strength is color-coded: blue connecting lines indicate strongly entangled orbital pairs, while purple lines denote weakly entangled orbital pairs.
$^3\Phi_{(\rm v)}$: $^3\Phi$ state for CUO$^{(\rm v)}$Ng$_4$. $^3\Phi_{(\rm a)}$: $^3\Phi$ state for CUO$^{(\rm a)}$Ng$_4$.
}\label{fig:diagrams}
\end{figure*}

\subsection{The CUO--Ng interaction dissected by orbital entanglement.}
Although, the complexation of the CUO molecule by the noble gas ligands is small, 
its effect on the spin-state splittings is remarkable and asks for an analysis of the quantum 
entanglement among the single-electron states.
Since, a spin-free wavefunction can be used to calculate the spin--orbit coupling to the first order of perturbation theory~\cite{CUO_Laura} in this particular case (\textit{cf}. Section ~\ref{sec:CUO}), it contains all the information necessary to study the entanglement in the unperturbed wavefunction. 
Therefore, the quantum information analysis of the spin-free DRMG wavefunction can be considered sufficient and reliable, although spin--orbit coupling gives the decisive energy contribution.

Fig.~\ref{fig:orbitals} shows the most important valence natural orbitals of CUONg$_4$ obtained at 
the equilibrium distances (cf. Table~\ref{tab:data}). These are the twelve highest occupied Ng$_4$ valence molecular orbitals (in Fig.~\ref{fig:orbitals}(a)) 
and the CUO $\sigma$-, $\pi$-, $\delta$- and $\phi$-molecular orbitals (in Fig.~\ref{fig:orbitals}(b)). We should note that the active orbitals 
are similar for CUONe$_4$ and CUOAr$_4$ and for all investigated spin states. Furthermore, the orbitals centered on the CUO unit do not differ from those of the bare CUO molecule. Surprisingly, even the U $5f_{\phi}$-orbital  remains unchanged in the $^3\Phi$ CUOAr$_4$ molecule. 
Due to the spatial similarity of the molecular orbitals obtained for different spin states and noble gas environments, an analysis based on an overlap measure between noble gas molecular orbitals and orbitals centered on the CUO moiety remains inconclusive and cannot explain the diverging stabilization energies in the Ne$_4$ and Ar$_4$ surrounding (the contribution of Ng$_4$ atomic orbitals to the CUO-centered molecular orbitals is negligible). 
Moreover, the examination of natural occupation numbers is less instructive since similar natural occupation numbers have been obtained in CASSCF and DMRG calculations where Ng$_4$ molecular orbitals remain doubly occupied along the dissociation pathway for all investigated spin states (\emph{cf.} Table ~I of the Supporting Information).

To elucidate the different complexation energies of the CUONg$_4$ compounds, different diagnostic tools are required that are not only based on occupation numbers and molecular orbital overlap measures. 
Recently, we have demonstrated \cite{entanglement_letter,entanglement_bonding_2013} that entanglement measures based on one- and two-orbital reduced density matrices represent a versatile tool for the analysis of electron correlation effects among molecular orbitals and facilitate a qualitative interpretation of electronic structures in terms of quantum correlation of  molecular orbitals \cite{book_chapter}. The mutual information\cite{legeza_dbss,Legeza2006,Rissler2006519} quantifies the interaction of each orbital pair $(i,j)$ embedded in all other orbitals of the active space 
(see Refs.~\citenum{entanglement_letter,entanglement_bonding_2013} for further details) and hence represents an adequate measure to assess the quantum entanglement of CUO and the noble gas surrounding directly from the electronic wave function. 
The mutual information is defined as
\begin{equation}
I_{i, j} = \frac{1}{2} ( s(2)_{i, j} - s(1)_i - s(1)_j)(1-\delta_{ij}),
\end{equation} 
where $s(1)_i  $ and $ s(2)_{i, j} $ are the one- and two-orbital entropy for orbital $i$ or orbital pair $(i,j)$, respectively, determined from the eigenvalues of the one- and two-orbital reduced density matrices \cite{entanglement_bonding_2013}, while $\delta_{ij}$ is the Kronecker delta.

The one- and two-orbital reduced density matrices are \emph{many}-particle reduced density matrices (up to 2 and 4 particles for $s(1)_i  $ and $ s(2)_{i, j} $, respectively), and hence contain more information than, \emph{e.g.}, the one-particle reduced density matrix, whose eigenvalues correspond to the natural occupation numbers.

The single-orbital entropy $s(1)$ is defined as
\begin{equation}
s(1)_i = - \sum_{\alpha} w_{\alpha,i} \ln w_{\alpha,i},
\label{eq:s1}
\end{equation}
where $w_{\alpha,i}$ is the eigenvalue of the one-orbital reduced density matrix of a given orbital \cite{entanglement_bonding_2013} ($\alpha$ denotes the four different occupations of a spatial orbital). The single-orbital entropy quantifies the entanglement between one particular orbital and the remaining set of orbitals contained in the active orbital space and can be used to dissect electron correlation effects in different contributions which can be distinguished with respect to their interaction strength.

\begin{figure*}[tb]
\centering
\includegraphics[width=0.55\linewidth]{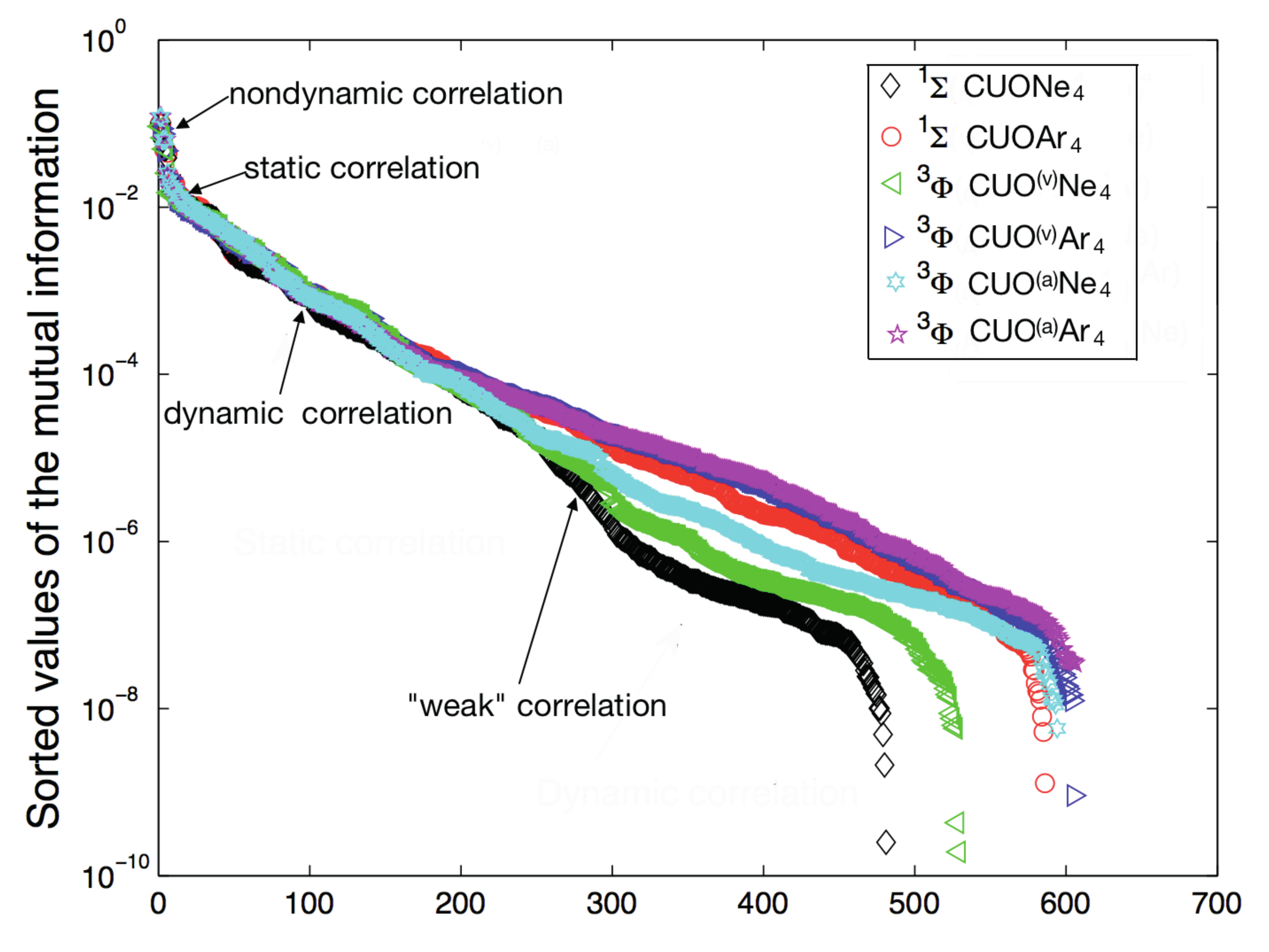}
\caption{
Decay of the mutual information for CUONe$_4$ and CUOAr$_4$ in different spin states. 
}\label{fig:s6}
\end{figure*}

\begin{figure*}[tb]
\centering
\includegraphics[width=0.95\linewidth]{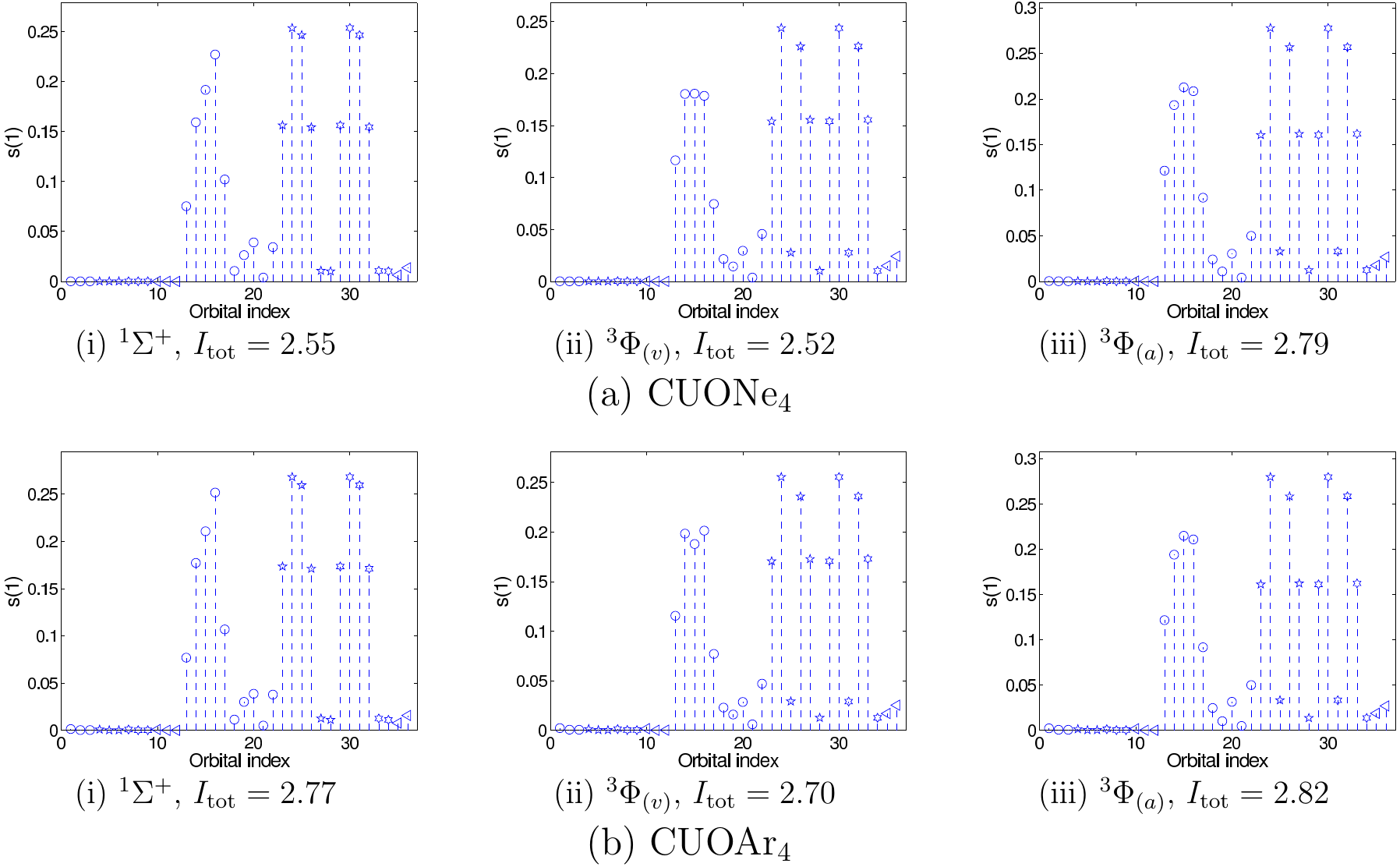}
\caption{
Single orbital entropy $s(1)$ for CUONg$_4$ determined at the equilibrium distance for Ng = Ne, Ar. The first twelve orbitals (starting from orbital index \#1) represent Ng$_4$ molecular orbitals, while the remaining molecular orbitals are centered on the CUO moiety. 
Orbital indices \#1, \#2, \#3, \#4, $\ldots$ of Fig. 5 correspond to orbital indices \#1, \#2, \#3, \#14, $\ldots$ of Fig. 4.
$^3\Phi_{(\rm v)}$: $^3\Phi$ state for CUO$^{(\rm v)}$Ng$_4$. $^3\Phi_{(\rm a)}$: $^3\Phi$ state for CUO$^{(\rm a)}$Ng$_4$.
}\label{fig:s1}
\end{figure*}

Fig.~\ref{fig:diagrams} displays the mutual information \cite{entanglement_letter,entanglement_bonding_2013} (lines are drawn if $I_{i, j}\geq 10^{-5}$) for all active orbital pairs in the CUONe$_4$ and CUOAr$_4$ molecules in their singlet and triplet states at their corresponding equilibrium structures determined from DMRG(38,36) wave functions. 
The entanglement diagrams for additional U--Ng distances (and different cut-off values for the mutual information) as well as the decay of the mutual information can be found in the Supporting Information.

In Fig.~\ref{fig:diagrams}, the interaction strength between the orbital pairs is color-coded (\emph{cf.} Ref. \citenum{entanglement_letter} for a detailed discussion): 
nondynamic electron correlation is indicated by blue (mutual information of order $\sim10^{-1}$) and static by red (mutual information of order $\sim10^{-2}$) connecting lines, while dynamic correlation is mainly attributed to orbitals connected by green (mutual information of order $\sim10^{-3}$) lines (see also Fig.~\ref{fig:s6} and the Supporting Information for a diagrammatic illustration). 
Note that the orbital index of Fig.~\ref{fig:diagrams} corresponds to the orbital number in Fig.~\ref{fig:orbitals} for the triplet state. For the singlet configuration, the orbital numbers are added in parentheses in Fig.~\ref{fig:orbitals} only if they differ from those of the triplet state.

The entanglement between the Ng$_4$ molecular orbitals and those molecular orbitals centered on the CUO unit is considerably weaker than between molecular orbital pairs centered on CUO only 
(solely purple connecting lines---mutual information of order $\sim10^{-5}$---for the former \emph{vs.} blue, red and green lines for the latter). This observation is in agreement with the weak nature of the CUO--Ng$_4$ interaction. In particular, the weakest interaction (\emph{i.e.}, the smallest number of purple lines between Ng$_4$ molecular orbitals and CUO-centered molecular orbitals) is found for $^1\Sigma^+$ CUONe$_4$, but gradually increases when going from $^3\Phi$ CUO$^{(\rm v)}$Ne$_4$ to $^3\Phi$ CUO$^{(\rm a)}$Ne$_4$. 
The differences in orbital entanglement are more clearly illustrated in Fig.~\ref{fig:s6} where the values of the mutual information are plotted in descending order for all investigated CUONg$_4$ compounds. While the decay of $I_{i,j}$ is similar for all CUONg$_4$ complexes if $I_{i,j} \ge 10^{-4}$, the evolution of the mutual information forks at $I_{i,j} \approx 10^{-4}$ (the forking regime was labeled as the weak correlation regime in Fig.~\ref{fig:s6}). Thus, different orbital entanglement patterns are obtained for small-valued $I_{i,j}$. All CUO complexes with argon atoms contain more weakly entangled orbitals ($I_{i,j} < 10^{-4}$) than CUO compounds in the Ne$_4$ surrounding. In addition, the decay of $I_{i,j}$ is in general faster for the $^1\Sigma$ state than for the $^3\Phi$ configuration of the CUO molecule. 

These entanglement patterns support the increasing potential well depth for $^1\Sigma^+$, over $^3\Phi$ 
CUO$^{(\rm v)}$Ne$_4$  to $^3\Phi$ CUO$^{(\rm a)}$Ne$_4$ as shown in Fig.~\ref{fig:PES-Ar}(a) and in Table~\ref{tab:data}. 
A qualitatively and quantitatively different entanglement picture is obtained for CUOAr$_4$, where a strong interaction between the CUO unit and the Ar$_4$ surrounding is already present for the $^1\Sigma^+$ state (note the large number of purple and turquoise lines between Ar$_4$ and CUO-centered molecular orbitals) and further increases when going from $^3\Phi$ 
CUO$^{(\rm v)}$Ar$_4$ to $^3\Phi$ CUO$^{(\rm a)}$Ar$_4$ (increasing number of purple connecting lines).

Since the interaction of the Ne$_4$ and Ar$_4$ surrounding with the CUO unit is very weak, the single orbital entropies of the noble gas molecular orbitals are close to Zero, while the single-orbital entropies corresponding to CUO-centered molecular orbitals are considerably larger (see Fig.~\ref{fig:s1}). 
Large values of the single-orbital entropy indicate that the electronic structure of the CUO unit is dominated by static electron correlation. 
However, we should note that the single-orbital entropy corresponding to the noble gas molecular orbitals are nevertheless larger for CUOAr$_4$ than for CUONe$_4$ (tiny differences are obtained for orbitals \#1--\#4 in Fig.~5) due to the stronger interaction of the CUO unit with the Ar$_4$ surrounding.
In addition, the total quantum information $I_{\rm tot}$ (the sum over single-orbital entropies \cite{entanglement_letter}, summarized in Fig. 5) is strictly larger for CUOAr$_4$ for all spin states. 
Since the structure of the CUO unit is similar for both CUONe$_4$ and CUOAr$_4$, these discrepancies can only be related to the change in the noble gas surrounding and thus support that the orbital entanglement between Ng$_4$ and CUO is stronger for Ar$_4$ than for Ne$_4$.

To conclude, Fig.~\ref{fig:s1} clearly demonstrates that the electronic structure of the CUO unit does not change upon noble gas variation from neon to argon in a given spin state (see Fig. 3(a)), while Figs.~\ref{fig:diagrams} and~\ref{fig:s6} illustrate the stronger interaction of CUO embedded in Ar$_4$ compared to the Ne$_4$ environment. 

\section{Conclusions}
In this work, we presented the first DMRG study of actinide chemistry. 
In particular, we investigated the electronic structure of the CUONe$_4$ and CUOAr$_4$ complexes and analyzed the quantum correlation between the central CUO moiety 
and the noble gas environment by means of orbital entanglement. 
The complexation of the CUO molecule by noble gases lowers the first excited $^3\Phi$ 
state with respect to the $^1\Sigma^+$ state compared to the bare CUO complex, whose ground state is a 
$^1\Sigma^+$ state. 
In general, the largest coordination energy is found for the $^3\Phi$ state for all noble gas matrices studied.
While in CUONe$_4$ the Ne$_4$ valence orbitals are only weakly entangled with molecular orbitals 
centered on CUO, CUOAr$_4$ features strongly entangled CUO--Ar$_4$ molecular orbital pairs, which promotes the stabilization of the $^3\Phi$ state of CUOAr$_4$ compared to its singlet state resulting in energetically equivalent spin states. 
With addition of spin-orbit coupling the energy difference between the CUO moiety embedded in neon and argon atoms is brought down to 0.02 eV, and therefore the anticipated ground-state spin crossover might occur.
 
Our entanglement study using mutual information points to different quantum correlations of the weakly coordinating noble-gas atoms by which the "mysterious" interaction of CUO with Ne$_4$ and Ar$_4$ can be explained. 
In particular, the total quantum information $I_{\rm tot}$ comprised in the CUONe$_4$ and  CUOAr$_4$ molecules indicates larger quantum entanglement of the Ar$_4$ orbitals with the CUO-centered molecular orbitals compared to the Ne$_4$ environment, although the difference in complexation energies is very small.   

\section*{Acknowledgement}
The authors gratefully acknowledge financial support from the Swiss national science foundation 
SNF (project 200020-144458/1) 
and from the Hungarian Research Fund (OTKA) under Grant No.~NN110360 and K100908.
{\"O}.L.\ acknowledges support from the Alexander von Humboldt foundation and from ETH Z\"urich
during his time as a visiting professor.





\begin{mcitethebibliography}{73}
\providecommand*{\natexlab}[1]{#1}
\providecommand*{\mciteSetBstSublistMode}[1]{}
\providecommand*{\mciteSetBstMaxWidthForm}[2]{}
\providecommand*{\mciteBstWouldAddEndPuncttrue}
  {\def\EndOfBibitem{\unskip.}}
\providecommand*{\mciteBstWouldAddEndPunctfalse}
  {\let\EndOfBibitem\relax}
\providecommand*{\mciteSetBstMidEndSepPunct}[3]{}
\providecommand*{\mciteSetBstSublistLabelBeginEnd}[3]{}
\providecommand*{\EndOfBibitem}{}
\mciteSetBstSublistMode{f}
\mciteSetBstMaxWidthForm{subitem}
{(\emph{\alph{mcitesubitemcount}})}
\mciteSetBstSublistLabelBeginEnd{\mcitemaxwidthsubitemform\space}
{\relax}{\relax}

\bibitem[Pepper and Bursten(1991)]{bursten_91}
M.~Pepper and B.~E. Bursten, \emph{Chem. Rev.}, 1991, \textbf{91},
  719--741\relax
\mciteBstWouldAddEndPuncttrue
\mciteSetBstMidEndSepPunct{\mcitedefaultmidpunct}
{\mcitedefaultendpunct}{\mcitedefaultseppunct}\relax
\EndOfBibitem
\bibitem[Denning(2007)]{denning07}
R.~G. Denning, \emph{J. Phys. Chem. A}, 2007, \textbf{111}, 4125--4143\relax
\mciteBstWouldAddEndPuncttrue
\mciteSetBstMidEndSepPunct{\mcitedefaultmidpunct}
{\mcitedefaultendpunct}{\mcitedefaultseppunct}\relax
\EndOfBibitem
\bibitem[Morss \emph{et~al.}(2010)Morss, Edelstein, and Fuger]{Actinides_bible}
L.~R. Morss, N.~M. Edelstein and J.~Fuger, \emph{The Chemistry of the Actinide
  and Transactinide Elements}, Springer, 2010\relax
\mciteBstWouldAddEndPuncttrue
\mciteSetBstMidEndSepPunct{\mcitedefaultmidpunct}
{\mcitedefaultendpunct}{\mcitedefaultseppunct}\relax
\EndOfBibitem
\bibitem[Burns \emph{et~al.}(2010)Burns, Ikeda, and Czerwinski]{MRS_bulletin}
P.~C. Burns, Y.~Ikeda and K.~Czerwinski, \emph{MRS Bulletin}, 2010,
  \textbf{35}, 868--876\relax
\mciteBstWouldAddEndPuncttrue
\mciteSetBstMidEndSepPunct{\mcitedefaultmidpunct}
{\mcitedefaultendpunct}{\mcitedefaultseppunct}\relax
\EndOfBibitem
\bibitem[Nash(1993)]{PUREX1}
K.~Nash, \emph{Solvent~Extr.~Ion~ Exch.}, 1993, \textbf{11}, 729--768\relax
\mciteBstWouldAddEndPuncttrue
\mciteSetBstMidEndSepPunct{\mcitedefaultmidpunct}
{\mcitedefaultendpunct}{\mcitedefaultseppunct}\relax
\EndOfBibitem
\bibitem[Sood and Patil(1995)]{sood95}
D.~D. Sood and S.~Patil, \emph{J.~Radioanyl.~Nuc.~Chem.}, 1995, \textbf{203},
  547--873\relax
\mciteBstWouldAddEndPuncttrue
\mciteSetBstMidEndSepPunct{\mcitedefaultmidpunct}
{\mcitedefaultendpunct}{\mcitedefaultseppunct}\relax
\EndOfBibitem
\bibitem[Nash \emph{et~al.}(2000)Nash, Barrans, Chiarizia, Dietz, Jensen,
  Rickert, Moyer, Bonnesen, Bryan, and Sachleben]{PUREX2}
K.~L. Nash, R.~E. Barrans, R.~Chiarizia, M.~L. Dietz, M.~Jensen, P.~Rickert,
  B.~A. Moyer, P.~V. Bonnesen, J.~C. Bryan and R.~A. Sachleben,
  \emph{Solvent~Extr.~Ion~ Exch.}, 2000, \textbf{18}, 605--631\relax
\mciteBstWouldAddEndPuncttrue
\mciteSetBstMidEndSepPunct{\mcitedefaultmidpunct}
{\mcitedefaultendpunct}{\mcitedefaultseppunct}\relax
\EndOfBibitem
\bibitem[Zhang \emph{et~al.}(2002)Zhang, Konduru, Dai, and
  Overbury]{U_catalysis_cc}
Z.~T. Zhang, M.~Konduru, S.~Dai and S.~H. Overbury, \emph{Chem. Comm.}, 2002,
  \textbf{20}, 2406--2407\relax
\mciteBstWouldAddEndPuncttrue
\mciteSetBstMidEndSepPunct{\mcitedefaultmidpunct}
{\mcitedefaultendpunct}{\mcitedefaultseppunct}\relax
\EndOfBibitem
\bibitem[Hutchings \emph{et~al.}(1996)Hutchings, Heneghan, Hudson, and
  Taylor]{U_catalysis_n}
G.~J. Hutchings, C.~S. Heneghan, I.~D. Hudson and S.~H. Taylor, \emph{Nature},
  1996, \textbf{384}, 341--343\relax
\mciteBstWouldAddEndPuncttrue
\mciteSetBstMidEndSepPunct{\mcitedefaultmidpunct}
{\mcitedefaultendpunct}{\mcitedefaultseppunct}\relax
\EndOfBibitem
\bibitem[Fox \emph{et~al.}(2008)Fox, Bart, Meyer, and
  Cummins]{Meyer_nature_2008}
A.~R. Fox, S.~C. Bart, K.~Meyer and C.~C. Cummins, \emph{Nature}, 2008,
  \textbf{455}, 341--349\relax
\mciteBstWouldAddEndPuncttrue
\mciteSetBstMidEndSepPunct{\mcitedefaultmidpunct}
{\mcitedefaultendpunct}{\mcitedefaultseppunct}\relax
\EndOfBibitem
\bibitem[Bart and Meyer(2008)]{Meyer_book_2008}
S.~C. Bart and K.~Meyer, \emph{{Organometallic and Coordination Chemistry of
  the Actinides}}, Springer, 2008, vol. 127, ch. {Highlights in Uranium
  Coordination Chemistry}, pp. 119--176\relax
\mciteBstWouldAddEndPuncttrue
\mciteSetBstMidEndSepPunct{\mcitedefaultmidpunct}
{\mcitedefaultendpunct}{\mcitedefaultseppunct}\relax
\EndOfBibitem
\bibitem[Castro-Rodriguez \emph{et~al.}(2004)Castro-Rodriguez, Nakai, Zakharov,
  Rheingold, and Meyer]{Castro-Rodriguez2004}
I.~Castro-Rodriguez, H.~Nakai, L.~N. Zakharov, A.~L. Rheingold and K.~Meyer,
  \emph{Science}, 2004, \textbf{305}, 1757--1759\relax
\mciteBstWouldAddEndPuncttrue
\mciteSetBstMidEndSepPunct{\mcitedefaultmidpunct}
{\mcitedefaultendpunct}{\mcitedefaultseppunct}\relax
\EndOfBibitem
\bibitem[Hayton \emph{et~al.}(2005)Hayton, Boncella, Scott, Palmer, Batista,
  and Hay]{Hayton2005}
T.~W. Hayton, J.~M. Boncella, B.~L. Scott, P.~D. Palmer, E.~R. Batista and
  P.~J. Hay, \emph{Science}, 2005, \textbf{310}, 1941--1943\relax
\mciteBstWouldAddEndPuncttrue
\mciteSetBstMidEndSepPunct{\mcitedefaultmidpunct}
{\mcitedefaultendpunct}{\mcitedefaultseppunct}\relax
\EndOfBibitem
\bibitem[Groenewold \emph{et~al.}(2008)Groenewold, {Van Stipdonk}, de~Jong,
  Oomens, Gresham, McIlwain, Gao, Siboulet, Visscher, Kullman, and
  Polfer]{Gary_2008}
G.~S. Groenewold, M.~J. {Van Stipdonk}, W.~A. de~Jong, J.~Oomens, G.~L.
  Gresham, M.~E. McIlwain, D.~Gao, B.~Siboulet, L.~Visscher, M.~Kullman and
  N.~Polfer, \emph{ChemPhysChem}, 2008, \textbf{9}, 1278--1285\relax
\mciteBstWouldAddEndPuncttrue
\mciteSetBstMidEndSepPunct{\mcitedefaultmidpunct}
{\mcitedefaultendpunct}{\mcitedefaultseppunct}\relax
\EndOfBibitem
\bibitem[Groenewold \emph{et~al.}(2010)Groenewold, van Stipdonk, Oomens,
  de~Jong, Gresham, and McIlwain]{Groenewold_2010}
G.~S. Groenewold, M.~J. van Stipdonk, J.~Oomens, W.~A. de~Jong, G.~L. Gresham
  and M.~E. McIlwain, \emph{Int. J. Mass Spectrom.}, 2010, \textbf{297},
  67--75\relax
\mciteBstWouldAddEndPuncttrue
\mciteSetBstMidEndSepPunct{\mcitedefaultmidpunct}
{\mcitedefaultendpunct}{\mcitedefaultseppunct}\relax
\EndOfBibitem
\bibitem[Hu \emph{et~al.}(2012)Hu, Qiu, Xiong, Schwarz, and Li]{CUO_Chem_Sci}
H.-S. Hu, Y.-H. Qiu, X.-G. Xiong, W.~H.~E. Schwarz and J.~Li, \emph{Chem.
  Sci.}, 2012, \textbf{3}, 2786--2796\relax
\mciteBstWouldAddEndPuncttrue
\mciteSetBstMidEndSepPunct{\mcitedefaultmidpunct}
{\mcitedefaultendpunct}{\mcitedefaultseppunct}\relax
\EndOfBibitem
\bibitem[Gagliardi and Roos(2005)]{Roos_U2}
L.~Gagliardi and B.~Roos, \emph{Nature}, 2005, \textbf{433}, 848--851\relax
\mciteBstWouldAddEndPuncttrue
\mciteSetBstMidEndSepPunct{\mcitedefaultmidpunct}
{\mcitedefaultendpunct}{\mcitedefaultseppunct}\relax
\EndOfBibitem
\bibitem[Wang \emph{et~al.}(2004)Wang, Andrews, Li, and Bursten]{wang_2004}
X.~Wang, L.~Andrews, J.~Li and B.~E. Bursten, \emph{Angew.~Chem.~Int.~Ed.},
  2004, \textbf{43}, 2554--2557\relax
\mciteBstWouldAddEndPuncttrue
\mciteSetBstMidEndSepPunct{\mcitedefaultmidpunct}
{\mcitedefaultendpunct}{\mcitedefaultseppunct}\relax
\EndOfBibitem
\bibitem[Lue \emph{et~al.}(2004)Lue, Jin, Ortiz, Rienstra-Kiracofe, and
  Heaven]{Heaven_UO2_04}
C.~J. Lue, J.~Jin, M.~J. Ortiz, J.~C. Rienstra-Kiracofe and M.~C. Heaven,
  \emph{J. Am. Chem. Soc.}, 2004, \textbf{126}, 1812--1815\relax
\mciteBstWouldAddEndPuncttrue
\mciteSetBstMidEndSepPunct{\mcitedefaultmidpunct}
{\mcitedefaultendpunct}{\mcitedefaultseppunct}\relax
\EndOfBibitem
\bibitem[Infante \emph{et~al.}(2010)Infante, Andrews, Wang, and
  Gagliardi]{ivan_2010}
I.~Infante, L.~Andrews, X.~Wang and L.~Gagliardi, \emph{Chem. Eur. J.}, 2010,
  \textbf{43}, 12804--12807\relax
\mciteBstWouldAddEndPuncttrue
\mciteSetBstMidEndSepPunct{\mcitedefaultmidpunct}
{\mcitedefaultendpunct}{\mcitedefaultseppunct}\relax
\EndOfBibitem
\bibitem[Tague \emph{et~al.}(1993)Tague, Andrews, and Hunt]{CUO_93}
T.~J. Tague, L.~Andrews and R.~D. Hunt, \emph{J. Phys. Chem.}, 1993,
  \textbf{97}, 10920--10924\relax
\mciteBstWouldAddEndPuncttrue
\mciteSetBstMidEndSepPunct{\mcitedefaultmidpunct}
{\mcitedefaultendpunct}{\mcitedefaultseppunct}\relax
\EndOfBibitem
\bibitem[Li \emph{et~al.}(2002)Li, Bursten, Liang, and Andrews]{CUO_Nb_science}
J.~Li, B.~E. Bursten, B.~Liang and L.~Andrews, \emph{Science}, 2002,
  \textbf{295}, 2242--2245\relax
\mciteBstWouldAddEndPuncttrue
\mciteSetBstMidEndSepPunct{\mcitedefaultmidpunct}
{\mcitedefaultendpunct}{\mcitedefaultseppunct}\relax
\EndOfBibitem
\bibitem[Liang \emph{et~al.}(2002)Liang, Andrews, Li, and Bursten]{CUO_Nb_JACS}
B.~Liang, L.~Andrews, J.~Li and B.~E. Bursten, \emph{J. Am. Chem. Soc.}, 2002,
  \textbf{124}, 9016--9017\relax
\mciteBstWouldAddEndPuncttrue
\mciteSetBstMidEndSepPunct{\mcitedefaultmidpunct}
{\mcitedefaultendpunct}{\mcitedefaultseppunct}\relax
\EndOfBibitem
\bibitem[Liang \emph{et~al.}(2003)Liang, Andrews, Li, and Bursten]{CUO_CEJ}
B.~Liang, L.~Andrews, J.~Li and B.~E. Bursten, \emph{Chem. Eur. J.}, 2003,
  \textbf{9}, 4781--4788\relax
\mciteBstWouldAddEndPuncttrue
\mciteSetBstMidEndSepPunct{\mcitedefaultmidpunct}
{\mcitedefaultendpunct}{\mcitedefaultseppunct}\relax
\EndOfBibitem
\bibitem[Andrews \emph{et~al.}(2003)Andrews, Liang, Li, and
  Bursten]{CUO_Nb_JACSa}
L.~Andrews, B.~Liang, J.~Li and B.~E. Bursten, \emph{J. Am. Chem. Soc.}, 2003,
  \textbf{125}, 3126--3139\relax
\mciteBstWouldAddEndPuncttrue
\mciteSetBstMidEndSepPunct{\mcitedefaultmidpunct}
{\mcitedefaultendpunct}{\mcitedefaultseppunct}\relax
\EndOfBibitem
\bibitem[Liang \emph{et~al.}(2004)Liang, Andrews, Li, and Bursten]{CUO_Nb_IC}
B.~Liang, L.~Andrews, J.~Li and B.~E. Bursten, \emph{Inorg. Chem.}, 2004,
  \textbf{43}, 882--894\relax
\mciteBstWouldAddEndPuncttrue
\mciteSetBstMidEndSepPunct{\mcitedefaultmidpunct}
{\mcitedefaultendpunct}{\mcitedefaultseppunct}\relax
\EndOfBibitem
\bibitem[Fleig(2011)]{Timo_overview}
T.~Fleig, \emph{Chem. Phys.}, 2011, \textbf{395}, 2--15\relax
\mciteBstWouldAddEndPuncttrue
\mciteSetBstMidEndSepPunct{\mcitedefaultmidpunct}
{\mcitedefaultendpunct}{\mcitedefaultseppunct}\relax
\EndOfBibitem
\bibitem[Saue(2012)]{Trond_rev_2012}
T.~Saue, \emph{Chem.~Phys.~Chem}, 2012, \textbf{3}, 3077--3094\relax
\mciteBstWouldAddEndPuncttrue
\mciteSetBstMidEndSepPunct{\mcitedefaultmidpunct}
{\mcitedefaultendpunct}{\mcitedefaultseppunct}\relax
\EndOfBibitem
\bibitem[Gagliardi and Roos(2000)]{CUO_2000}
L.~Gagliardi and B.~O. Roos, \emph{Chem. Phys. Lett.}, 2000, \textbf{331},
  229\relax
\mciteBstWouldAddEndPuncttrue
\mciteSetBstMidEndSepPunct{\mcitedefaultmidpunct}
{\mcitedefaultendpunct}{\mcitedefaultseppunct}\relax
\EndOfBibitem
\bibitem[Bursten \emph{et~al.}(2003)Bursten, Drummond, and Li]{Bursten_2003}
B.~E. Bursten, M.~L. Drummond and J.~Li, \emph{Faraday Discuss.}, 2003,
  \textbf{124}, 1--24\relax
\mciteBstWouldAddEndPuncttrue
\mciteSetBstMidEndSepPunct{\mcitedefaultmidpunct}
{\mcitedefaultendpunct}{\mcitedefaultseppunct}\relax
\EndOfBibitem
\bibitem[Roos \emph{et~al.}(2003)Roos, P.-O.Widmark, and Gagliardi]{CUO_Laura}
B.~O. Roos, P.-O.Widmark and L.~Gagliardi, \emph{Faraday Discuss.}, 2003,
  \textbf{124}, 57--62\relax
\mciteBstWouldAddEndPuncttrue
\mciteSetBstMidEndSepPunct{\mcitedefaultmidpunct}
{\mcitedefaultendpunct}{\mcitedefaultseppunct}\relax
\EndOfBibitem
\bibitem[Infante and Visscher(2004)]{CUO_Ivan}
I.~Infante and L.~Visscher, \emph{J. Chem. Phys.}, 2004, \textbf{121},
  5783--5788\relax
\mciteBstWouldAddEndPuncttrue
\mciteSetBstMidEndSepPunct{\mcitedefaultmidpunct}
{\mcitedefaultendpunct}{\mcitedefaultseppunct}\relax
\EndOfBibitem
\bibitem[Yang \emph{et~al.}(2009)Yang, Tyagi, Zhang, and Pitzer]{CUO_CI}
T.~Yang, R.~Tyagi, Z.~Zhang and R.~M. Pitzer, \emph{Mol. Phys.}, 2009,
  \textbf{107}, 1193--1195\relax
\mciteBstWouldAddEndPuncttrue
\mciteSetBstMidEndSepPunct{\mcitedefaultmidpunct}
{\mcitedefaultendpunct}{\mcitedefaultseppunct}\relax
\EndOfBibitem
\bibitem[Tecmer \emph{et~al.}(2012)Tecmer, van Lingen, Gomes, and
  Visscher]{CUO_pawel}
P.~Tecmer, H.~van Lingen, A.~S.~P. Gomes and L.~Visscher, \emph{J. Chem.
  Phys.}, 2012, \textbf{137}, 084308\relax
\mciteBstWouldAddEndPuncttrue
\mciteSetBstMidEndSepPunct{\mcitedefaultmidpunct}
{\mcitedefaultendpunct}{\mcitedefaultseppunct}\relax
\EndOfBibitem
\bibitem[Wang \emph{et~al.}(2012)Wang, van Gunsteren, and
  Chai]{actinoid_rev_2012}
D.~Wang, W.~F. van Gunsteren and Z.~Chai, \emph{Chem. Soc. Rev.}, 2012,
  \textbf{41}, 5836--5865\relax
\mciteBstWouldAddEndPuncttrue
\mciteSetBstMidEndSepPunct{\mcitedefaultmidpunct}
{\mcitedefaultendpunct}{\mcitedefaultseppunct}\relax
\EndOfBibitem
\bibitem[Tecmer \emph{et~al.}(2013)Tecmer, Govind, Kowalski, de~Jong., and
  Visscher]{pawel_saldien}
P.~Tecmer, N.~Govind, K.~Kowalski, W.~A. de~Jong. and L.~Visscher, \emph{J.
  Chem. Phys.}, 2013, \textbf{139}, 034301\relax
\mciteBstWouldAddEndPuncttrue
\mciteSetBstMidEndSepPunct{\mcitedefaultmidpunct}
{\mcitedefaultendpunct}{\mcitedefaultseppunct}\relax
\EndOfBibitem
\bibitem[Malmqvist \emph{et~al.}(2002)Malmqvist, Roos, and
  Schimmelpfennig]{Malmqvist2002}
P.~A. Malmqvist, B.~O. Roos and B.~Schimmelpfennig, \emph{Chem. Phys.
  Lett.}, 2002, \textbf{357}, 230--240\relax
\mciteBstWouldAddEndPuncttrue
\mciteSetBstMidEndSepPunct{\mcitedefaultmidpunct}
{\mcitedefaultendpunct}{\mcitedefaultseppunct}\relax
\EndOfBibitem
\bibitem[Werner and Knowles(1985)]{Werner_1985}
H.-J. Werner and P.~J. Knowles, \emph{J. Chem. Phys.}, 1985, \textbf{82},
  5053--5063\relax
\mciteBstWouldAddEndPuncttrue
\mciteSetBstMidEndSepPunct{\mcitedefaultmidpunct}
{\mcitedefaultendpunct}{\mcitedefaultseppunct}\relax
\EndOfBibitem
\bibitem[Schollw\"ock(2005)]{scholl05}
U.~Schollw\"ock, \emph{Rev. Mod. Phys.}, 2005, \textbf{77}, 259--315\relax
\mciteBstWouldAddEndPuncttrue
\mciteSetBstMidEndSepPunct{\mcitedefaultmidpunct}
{\mcitedefaultendpunct}{\mcitedefaultseppunct}\relax
\EndOfBibitem
\bibitem[White(1992)]{white}
S.~R. White, \emph{Phys. Rev. Lett.}, 1992, \textbf{69}, 2863--2866\relax
\mciteBstWouldAddEndPuncttrue
\mciteSetBstMidEndSepPunct{\mcitedefaultmidpunct}
{\mcitedefaultendpunct}{\mcitedefaultseppunct}\relax
\EndOfBibitem
\bibitem[Legeza \emph{et~al.}(2008)Legeza, Noack, S\'olyom, and
  Tincani]{ors_springer}
O.~Legeza, R.~Noack, J.~S\'olyom and L.~Tincani, \emph{Computational
  Many-Particle Physics}, Springer, Berlin/Heidelerg, 2008, vol. 739, pp.
  653--664\relax
\mciteBstWouldAddEndPuncttrue
\mciteSetBstMidEndSepPunct{\mcitedefaultmidpunct}
{\mcitedefaultendpunct}{\mcitedefaultseppunct}\relax
\EndOfBibitem
\bibitem[Marti and Reiher(2010)]{marti2010b}
K.~H. Marti and M.~Reiher, \emph{Z. Phys. Chem.}, 2010, \textbf{224},
  583--599\relax
\mciteBstWouldAddEndPuncttrue
\mciteSetBstMidEndSepPunct{\mcitedefaultmidpunct}
{\mcitedefaultendpunct}{\mcitedefaultseppunct}\relax
\EndOfBibitem
\bibitem[Chan and Sharma(2011)]{chanreview}
G.~K.-L. Chan and S.~Sharma, \emph{Annu. Rev. Phys. Chem.}, 2011, \textbf{62},
  465--481\relax
\mciteBstWouldAddEndPuncttrue
\mciteSetBstMidEndSepPunct{\mcitedefaultmidpunct}
{\mcitedefaultendpunct}{\mcitedefaultseppunct}\relax
\EndOfBibitem
\bibitem[Marti \emph{et~al.}(2008)Marti, Ondik, Moritz, and Reiher]{marti2008}
K.~H. Marti, I.~M. Ondik, G.~Moritz and M.~Reiher, \emph{J. Chem. Phys.}, 2008,
  \textbf{128}, 014104\relax
\mciteBstWouldAddEndPuncttrue
\mciteSetBstMidEndSepPunct{\mcitedefaultmidpunct}
{\mcitedefaultendpunct}{\mcitedefaultseppunct}\relax
\EndOfBibitem
\bibitem[Boguslawski \emph{et~al.}(2012)Boguslawski, Marti, Legeza, and
  Reiher]{fenoDMRG}
K.~Boguslawski, K.~H. Marti, O.~Legeza and M.~Reiher, \emph{J. Chem. Theory
  Comput.}, 2012, \textbf{8}, 1970--1982\relax
\mciteBstWouldAddEndPuncttrue
\mciteSetBstMidEndSepPunct{\mcitedefaultmidpunct}
{\mcitedefaultendpunct}{\mcitedefaultseppunct}\relax
\EndOfBibitem
\bibitem[Kurashige \emph{et~al.}(2013)Kurashige, Chan, and
  Yanai]{kurashige2013}
Y.~Kurashige, G.~K.-L. Chan and T.~Yanai, \emph{Nature Chem.}, 2013,
 \textbf{5}, 660--666 \relax
\mciteBstWouldAddEndPuncttrue
\mciteSetBstMidEndSepPunct{\mcitedefaultmidpunct}
{\mcitedefaultendpunct}{\mcitedefaultseppunct}\relax
\EndOfBibitem
\bibitem[Boguslawski \emph{et~al.}(2012)Boguslawski, Tecmer, Legeza, and
  Reiher]{entanglement_letter}
K.~Boguslawski, P.~Tecmer, O.~Legeza and M.~Reiher, \emph{J. Phys. Chem.
  Lett.}, 2012, \textbf{3}, 3129--3135\relax
\mciteBstWouldAddEndPuncttrue
\mciteSetBstMidEndSepPunct{\mcitedefaultmidpunct}
{\mcitedefaultendpunct}{\mcitedefaultseppunct}\relax
\EndOfBibitem
\bibitem[Barcza \emph{et~al.}(2011)Barcza, Legeza, Marti, and
  Reiher]{orbitalordering}
G.~Barcza, O.~Legeza, K.~H. Marti and M.~Reiher, \emph{Phys. Rev. A}, 2011,
  \textbf{83}, 012508\relax
\mciteBstWouldAddEndPuncttrue
\mciteSetBstMidEndSepPunct{\mcitedefaultmidpunct}
{\mcitedefaultendpunct}{\mcitedefaultseppunct}\relax
\EndOfBibitem
\bibitem[Boguslawski \emph{et~al.}(2013)Boguslawski, Tecmer, Barcza, Legeza,
  and Reiher]{entanglement_bonding_2013}
K.~Boguslawski, P.~Tecmer, G.~Barcza, O.~Legeza and M.~Reiher, \emph{J. Chem.
  Theory Comput.}, 2013, \textbf{9}, 2959--2973\relax
\mciteBstWouldAddEndPuncttrue
\mciteSetBstMidEndSepPunct{\mcitedefaultmidpunct}
{\mcitedefaultendpunct}{\mcitedefaultseppunct}\relax
\EndOfBibitem
\bibitem[Roos \emph{et~al.}(2004)Roos, Lindh, Malmqvist, Veryazov, and
  Widmark]{ANO-RCC_main_group}
B.~O. Roos, R.~Lindh, P.-A. Malmqvist, V.~Veryazov and P.-O. Widmark, \emph{J.
  Phys. Chem. A}, 2004, \textbf{108}, 2851--2858\relax
\mciteBstWouldAddEndPuncttrue
\mciteSetBstMidEndSepPunct{\mcitedefaultmidpunct}
{\mcitedefaultendpunct}{\mcitedefaultseppunct}\relax
\EndOfBibitem
\bibitem[Roos \emph{et~al.}(2005)Roos, Lindh, Malmqvist, Veryazov, and
  Widmark]{ANO-RCC_U}
B.~O. Roos, R.~Lindh, P.-A. Malmqvist, V.~Veryazov and P.-O. Widmark,
  \emph{Chem. Phys. Lett.}, 2005, \textbf{409}, 295--299\relax
\mciteBstWouldAddEndPuncttrue
\mciteSetBstMidEndSepPunct{\mcitedefaultmidpunct}
{\mcitedefaultendpunct}{\mcitedefaultseppunct}\relax
\EndOfBibitem
\bibitem[Hess(1986)]{dkh2}
B.~A. Hess, \emph{Phys.~Rev.~A}, 1986, \textbf{33}, 3742\relax
\mciteBstWouldAddEndPuncttrue
\mciteSetBstMidEndSepPunct{\mcitedefaultmidpunct}
{\mcitedefaultendpunct}{\mcitedefaultseppunct}\relax
\EndOfBibitem
\bibitem[Reiher and Wolf(2004)]{Reiher_2004a}
M.~Reiher and A.~Wolf, \emph{J. Chem. Phys.}, 2004, \textbf{121},
  2037--2047\relax
\mciteBstWouldAddEndPuncttrue
\mciteSetBstMidEndSepPunct{\mcitedefaultmidpunct}
{\mcitedefaultendpunct}{\mcitedefaultseppunct}\relax
\EndOfBibitem
\bibitem[Reiher and Wolf(2004)]{Reiher_2004b}
M.~Reiher and A.~Wolf, \emph{J. Chem. Phys.}, 2004, \textbf{121},
  10945--10956\relax
\mciteBstWouldAddEndPuncttrue
\mciteSetBstMidEndSepPunct{\mcitedefaultmidpunct}
{\mcitedefaultendpunct}{\mcitedefaultseppunct}\relax
\EndOfBibitem
\bibitem[{Werner} \emph{et~al.}(2010){Werner}, {Knowles}, Lindh, Manby,
  Sch\"utz, Celani, Korona, Mitrushenkov, Rauhut, Adler, and \emph{et
  al.}]{MOLPRO}
H.-J. {Werner}, P.~J. {Knowles}, R.~Lindh, F.~R. Manby, M.~Sch\"utz, P.~Celani,
  T.~Korona, A.~Mitrushenkov, G.~Rauhut, T.~B. Adler and \emph{et al.},
  \emph{{MOLPRO, Version 2010.1, a Package of \emph{Ab initio} Programs,
  Cardiff University: Cardiff, United Kingdom, and University of Stuttgart:
  Stuttgart, Germany}}, 2010\relax
\mciteBstWouldAddEndPuncttrue
\mciteSetBstMidEndSepPunct{\mcitedefaultmidpunct}
{\mcitedefaultendpunct}{\mcitedefaultseppunct}\relax
\EndOfBibitem
\bibitem[Pierloot and van Besien(2005)]{pierloot05}
K.~Pierloot and E.~van Besien, \emph{J. Chem. Phys.}, 2005, \textbf{123},
  204309\relax
\mciteBstWouldAddEndPuncttrue
\mciteSetBstMidEndSepPunct{\mcitedefaultmidpunct}
{\mcitedefaultendpunct}{\mcitedefaultseppunct}\relax
\EndOfBibitem
\bibitem[Legeza()]{dmrg_ors}
O.~Legeza, \emph{\textsc{QC-DMRG-Budapest}, A Program for Quantum Chemical
  {DMRG} Calculations. { \rm Copyright 2000--2013, HAS RISSPO Budapest}}\relax
\mciteBstWouldAddEndPuncttrue
\mciteSetBstMidEndSepPunct{\mcitedefaultmidpunct}
{\mcitedefaultendpunct}{\mcitedefaultseppunct}\relax
\EndOfBibitem
\bibitem[Legeza and S\'olyom(2003)]{legeza_dbss}
O.~Legeza and J.~S\'olyom, \emph{Phys. Rev. B}, 2003, \textbf{68}, 195116\relax
\mciteBstWouldAddEndPuncttrue
\mciteSetBstMidEndSepPunct{\mcitedefaultmidpunct}
{\mcitedefaultendpunct}{\mcitedefaultseppunct}\relax
\EndOfBibitem
\bibitem[Legeza \emph{et~al.}(2003)Legeza, R\"oder, and Hess]{legeza_dbss2}
O.~Legeza, J.~R\"oder and B.~A. Hess, \emph{Phys. Rev. B}, 2003, \textbf{67},
  125114\relax
\mciteBstWouldAddEndPuncttrue
\mciteSetBstMidEndSepPunct{\mcitedefaultmidpunct}
{\mcitedefaultendpunct}{\mcitedefaultseppunct}\relax
\EndOfBibitem
\bibitem[Legeza and S\'olyom(2004)]{legeza_dbss3}
O.~Legeza and J.~S\'olyom, \emph{Phys. Rev. B}, 2004, \textbf{70}, 205118\relax
\mciteBstWouldAddEndPuncttrue
\mciteSetBstMidEndSepPunct{\mcitedefaultmidpunct}
{\mcitedefaultendpunct}{\mcitedefaultseppunct}\relax
\EndOfBibitem
\bibitem[Pyykk\"{o} \emph{et~al.}(1994)Pyykk\"{o}, Li, and Runeberg]{pyykko}
P.~Pyykk\"{o}, J.~Li and N.~Runeberg, \emph{J. Phys. Chem.}, 1994, \textbf{98},
  4809--4813\relax
\mciteBstWouldAddEndPuncttrue
\mciteSetBstMidEndSepPunct{\mcitedefaultmidpunct}
{\mcitedefaultendpunct}{\mcitedefaultseppunct}\relax
\EndOfBibitem
\bibitem[Kaltsoyannis(2000)]{kaltsoyannis}
N.~Kaltsoyannis, \emph{Inorg. Chem.}, 2000, \textbf{39}, 6009--6017\relax
\mciteBstWouldAddEndPuncttrue
\mciteSetBstMidEndSepPunct{\mcitedefaultmidpunct}
{\mcitedefaultendpunct}{\mcitedefaultseppunct}\relax
\EndOfBibitem
\bibitem[Matsika \emph{et~al.}(2001)Matsika, Zhang, Brozell, Blaudeau, Wang,
  and Pitzer]{matsika_2001}
S.~Matsika, Z.~Zhang, S.~R. Brozell, J.-P. Blaudeau, Q.~Wang and R.~M. Pitzer,
  \emph{J. Phys. Chem. A}, 2001, \textbf{105}, 3825--3828\relax
\mciteBstWouldAddEndPuncttrue
\mciteSetBstMidEndSepPunct{\mcitedefaultmidpunct}
{\mcitedefaultendpunct}{\mcitedefaultseppunct}\relax
\EndOfBibitem
\bibitem[Clavagu{\'e}ra-Sarrio \emph{et~al.}(2004)Clavagu{\'e}ra-Sarrio,
  Ismail, Marsden, B{\'e}gue, and Pouchan]{XUY}
C.~Clavagu{\'e}ra-Sarrio, N.~Ismail, C.~J. Marsden, D.~B{\'e}gue and
  C.~Pouchan, \emph{Chem. Phys.}, 2004, \textbf{302}, 1--11\relax
\mciteBstWouldAddEndPuncttrue
\mciteSetBstMidEndSepPunct{\mcitedefaultmidpunct}
{\mcitedefaultendpunct}{\mcitedefaultseppunct}\relax
\EndOfBibitem
\bibitem[Tecmer \emph{et~al.}(2011)Tecmer, Gomes, Ekstr\"om, and
  Visscher]{pawel1}
P.~Tecmer, A.~S.~P. Gomes, U.~Ekstr\"om and L.~Visscher, \emph{Phys. Chem.
  Chem. Phys.}, 2011, \textbf{13}, 6249--6259\relax
\mciteBstWouldAddEndPuncttrue
\mciteSetBstMidEndSepPunct{\mcitedefaultmidpunct}
{\mcitedefaultendpunct}{\mcitedefaultseppunct}\relax
\EndOfBibitem
\bibitem[Coxon(1992)]{Coxon_1992}
J.~A. Coxon, \emph{J. Mol. Spectrosc.}, 1992, \textbf{282}, 274--282\relax
\mciteBstWouldAddEndPuncttrue
\mciteSetBstMidEndSepPunct{\mcitedefaultmidpunct}
{\mcitedefaultendpunct}{\mcitedefaultseppunct}\relax
\EndOfBibitem
\bibitem[avo()]{avogadro_1}
Avogadro: an Open-Source Molecular Builder and Visualization Tool. Version
  1.1.0 http://avogadro.openmolecules.net/\relax
\mciteBstWouldAddEndPuncttrue
\mciteSetBstMidEndSepPunct{\mcitedefaultmidpunct}
{\mcitedefaultendpunct}{\mcitedefaultseppunct}\relax
\EndOfBibitem
\bibitem[Hanwell \emph{et~al.}(2012)Hanwell, Curtis, Lonie, Vandermeersch,
  Zurek, and Hutchison]{avogadro_2}
M.~D. Hanwell, D.~E. Curtis, D.~C. Lonie, T.~Vandermeersch, E.~Zurek and G.~R.
  Hutchison, \emph{J. Cheminf.}, 2012, \textbf{4}, 1--17\relax
\mciteBstWouldAddEndPuncttrue
\mciteSetBstMidEndSepPunct{\mcitedefaultmidpunct}
{\mcitedefaultendpunct}{\mcitedefaultseppunct}\relax
\EndOfBibitem
\bibitem[Reiher(2002)]{M1}
M.~Reiher, \emph{Inorg. Chem.}, 2002, \textbf{41}, 6928--6935\relax
\mciteBstWouldAddEndPuncttrue
\mciteSetBstMidEndSepPunct{\mcitedefaultmidpunct}
{\mcitedefaultendpunct}{\mcitedefaultseppunct}\relax
\EndOfBibitem
\bibitem[Paulsen \emph{et~al.}(2013)Paulsen, Sch\"unemann, and Wolny]{M2}
H.~Paulsen, V.~Sch\"unemann and J.~A. Wolny, \emph{Eur. J. Inorg. Chem.}, 2013,
  \textbf{2013}, 628--641\relax
\mciteBstWouldAddEndPuncttrue
\mciteSetBstMidEndSepPunct{\mcitedefaultmidpunct}
{\mcitedefaultendpunct}{\mcitedefaultseppunct}\relax
\EndOfBibitem
\bibitem[Boguslawski and Reiher(2014)]{book_chapter}
K.~Boguslawski and M.~Reiher, \emph{Chemical Bonding Across the Periodic
  Table}, Wiley-VCH, 2014, vol. 2 of 'The Nature of the Chemical Bond
  Revisited', ch. Chemical Bonding in Open-Shell Transition Metal
  Complexes\relax
\mciteBstWouldAddEndPuncttrue
\mciteSetBstMidEndSepPunct{\mcitedefaultmidpunct}
{\mcitedefaultendpunct}{\mcitedefaultseppunct}\relax
\EndOfBibitem
\bibitem[Legeza and S\'olyom(2006)]{Legeza2006}
O.~Legeza and J.~S\'olyom, \emph{Phys. Rev. Lett.}, 2006, \textbf{96},
  116401\relax
\mciteBstWouldAddEndPuncttrue
\mciteSetBstMidEndSepPunct{\mcitedefaultmidpunct}
{\mcitedefaultendpunct}{\mcitedefaultseppunct}\relax
\EndOfBibitem
\bibitem[Rissler \emph{et~al.}(2006)Rissler, Noack, and White]{Rissler2006519}
J.~Rissler, R.~M. Noack and S.~R. White, \emph{Chem. Phys.}, 2006,
  \textbf{323}, 519--531\relax
\mciteBstWouldAddEndPuncttrue
\mciteSetBstMidEndSepPunct{\mcitedefaultmidpunct}
{\mcitedefaultendpunct}{\mcitedefaultseppunct}\relax
\EndOfBibitem
\end{mcitethebibliography}
\providecommand*{\mcitethebibliography}{\thebibliography}
\csname @ifundefined\endcsname{endmcitethebibliography}
{\let\endmcitethebibliography\endthebibliography}{}

\end{document}